\documentclass[reprint,superscriptaddress,amsmath,amssymb,aps]{revtex4-1}

\usepackage{graphicx}
\usepackage{dcolumn}
\usepackage{bm}
\usepackage{subfigure}
\usepackage{natbib}
\usepackage{multirow}
\usepackage{soul}
\usepackage{float}
\usepackage[normalem]{ulem}
\usepackage[usenames,dvipsnames]{color}

\def\QE{\textsc{Quantum ESPRESSO}\,}

\begin{document}

\title{Electronic structure of pristine and Ni-substituted LaFeO$_3$ from near edge x-ray absorption fine structure experiments and first-principles simulations}

\author{Iurii Timrov}\email[e-mail:]{ iurii.timrov@epfl.ch}
\affiliation{Theory and Simulation of Materials (THEOS), and National Centre for Computational Design and Discovery of Novel Materials (MARVEL), \'Ecole Polytechnique F\'ed\'erale de Lausanne (EPFL), CH-1015 Lausanne, Switzerland}

\author{Piyush Agrawal}
\affiliation{nanotech@surfaces Laboratory, and National Centre for Computational Design and Discovery of Novel Materials (MARVEL), Empa--Swiss Federal Laboratories for Materials Science and Technology, CH-8600 D\"ubendorf, Switzerland}
\affiliation{Electron Microscopy Center, Empa--Swiss Federal Laboratories for Materials Science and Technology, CH-8600 D\"ubendorf, Switzerland}

\author{Xinyu Zhang}
\email[e-mail:]{ xyzhang@ysu.edu.cn}
\affiliation{nanotech@surfaces Laboratory, Empa--Swiss Federal Laboratories for Materials Science and Technology, CH-8600 D\"ubendorf, Switzerland}
\affiliation{State Key Laboratory of Metastable Material Science and Technology, Yanshan University, CN-066004 Qinhuangdao, China}

\author{Selma Erat}
\affiliation{High Performance Ceramics Laboratory, Empa--Swiss Federal Laboratories for Materials Science and Technology, CH-8600 Dubendorf, Switzerland}
\affiliation{Department for Materials, Nonmetallic Inorganic Materials, ETH Zurich, Swiss Federal Institute of Technology, CH-8037 Zurich, Switzerland}
\affiliation{Vocational School of Technical Sciences, Department of Medical Services and Techniques, Program of Opticianry, Mersin University, TR-33343 Yenisehir, Mersin, Turkey}
\affiliation{Advanced Technology Education Research and Application Center, Mersin University, TR-33343 Yenisehir, Mersin, Turkey}

\author{Riping Liu}
\affiliation{State Key Laboratory of Metastable Material Science and Technology, Yanshan University, CN-066004 Qinhuangdao, China}

\author{Artur Braun}
\affiliation{High Performance Ceramics Laboratory, Empa--Swiss Federal Laboratories for Materials Science and Technology, CH-8600 Dubendorf, Switzerland}

\author{Matteo Cococcioni}
\affiliation{Department of Physics, University of Pavia, via Bassi 6, I-27100 Pavia, Italy}

\author{Matteo Calandra}
\affiliation{Dipartmento di Fisica, Universit\`a di Trento, Via Sommarive 14, 38123 Povo, Italy}
\affiliation{Sorbonne Universit\'e, CNRS, Institut des Nanosciences de Paris, UMR7588, F-75252 Paris, France}

\author{Nicola Marzari}
\affiliation{Theory and Simulation of Materials (THEOS), and National Centre for Computational Design and Discovery of Novel Materials (MARVEL), \'Ecole Polytechnique F\'ed\'erale de Lausanne (EPFL), CH-1015 Lausanne, Switzerland}

\author{Daniele Passerone}
\affiliation{nanotech@surfaces Laboratory, and National Centre for Computational Design and Discovery of Novel Materials (MARVEL), Empa--Swiss Federal Laboratories for Materials Science and Technology, CH-8600 D\"ubendorf, Switzerland}

\date{\today}

\begin{abstract}
We present a joint theoretical and experimental study of the oxygen $K$-edge spectra for LaFeO$_3$ and homovalent Ni-substituted LaFeO$_3$ (LaFe$_{0.75}$Ni$_{0.25}$O$_3$), using first-principles simulations based on density-functional theory with extended Hubbard functionals and x-ray absorption near edge structure (XANES) measurements. Ground-state and excited-state XANES calculations employ Hubbard on-site $U$ and inter-site $V$ parameters determined from first principles and the Lanczos recursive method to obtain absorption cross sections, which allows for a reliable description of XANES spectra in transition-metal compounds in a very broad energy range, with an accuracy comparable to that of hybrid functionals but at a substantially lower cost. We show that standard gradient-corrected exchange-correlation functionals fail in capturing accurately the electronic properties of both materials. In particular, for LaFe$_{0.75}$Ni$_{0.25}$O$_3$ they do not reproduce its semiconducting behavior and provide a poor description of the pre-edge features at the O $K$ edge. The inclusion of Hubbard interactions leads to a drastic improvement, accounting for the semiconducting ground state of LaFe$_{0.75}$Ni$_{0.25}$O$_3$ and for good agreement between calculated and measured XANES spectra. We show that the partial substitution of Ni for Fe affects the conduction-band bottom by generating a strongly hybridized O($2p$)--Ni($3d$) minority-spin empty electronic state. The present work, based on a consistent correction of self-interaction errors, outlines the crucial role of extended Hubbard functionals to describe the electronic structure of complex transition-metal oxides such as LaFeO$_3$ and LaFe$_{0.75}$Ni$_{0.25}$O$_3$ and paves the way to future studies on similar systems. 
\end{abstract}

\maketitle

\section{Introduction}
\label{sec:Intro}

Transition-metal (TM) oxides have been the subject of extensive studies for more than 60 years due to the variety of unique physical properties that are of great scientific and technological interest, such as their use as electrodes in energy converters (e.g. solid oxide fuel cells and solid oxide electrolysis cells)~\cite{Jacobson:2010}, in energy storage~\cite{Poizot:2000, Mizushima:1980}, spintronics~\cite{Raquet:1998}, gas-sensing~\cite{Wollenstein:2003}, and photo-catalysis~\cite{Demkov:2014}.
Among them, ABO$_3$ perovskites have attracted special attention: they can exhibit e.g. metal-insulator phase transitions~\cite{Subedi:2015, Mercy:2017}, show interesting defect chemistry~\cite{Marthinsen:2016, Becher:2015}, or give rise to entire classes of ferroelectric~\cite{Kimura:2019} and magnetoelectric multiferroic materials~\cite{Weston:2016, Spaldin:2019}. Doping and substituting A and/or B sites can lead to substantial changes in the physical and chemical properties of these materials with controlled valence; one example relevant to this work is that of the LaSrMn-oxides and LaSrFe-oxides~\cite{Abbate:1992}. Obviously, both A- and B-site substitutions add more complexity to the system, and key issues are how the type and concentration of the A- and/or B-site dopants can affect the crystallographic or electronic structure, and, in turn, transport properties. 
%Here we present an experimental and computational study on the B-site subsituted LaSrFeO$_3$.

LaFeO$_3$ (LFO) and its A- and/or B-site substituted materials have attracted attention because of their high conductivity at intermediate (400$^\circ$ -- 600$^\circ$~C) temperatures~\cite{Yahiro:2007}. LFO belongs to the class of rare-earth orthoferrites: it is an antiferromagnetic (AFM) insulator with a N\'eel temperature of $T_\mathrm{N} = 750$~K~\cite{Hearne:1995, Falcon:1997} and it has Fe$^{3+}$ oxidation with a $3d^5$ high spin electronic configuration $t^3_{2g} e^2_g$ ($^6A_{1g}$)~\cite{Erat:2010b, Abbate:1992}. Quite recently some studies have focused on heterovalent A-site and homovalent B-site substitutions with the goal of relating various spectroscopic signatures with electrical conductivity~\cite{Braun:2009, Erat:2009, Erat:2010}. Since it is of great importance to investigate the correlation between electronic structure and electronic conductivity, a systematic study of the hole doping states of La$_{1-x}$Sr$_x$Fe$_{0.75}$Ni$_{0.25}$O$_{3-\delta}$ was carried out by synchrotron x-ray spectroscopy and ligand-field multiplet calculations~\cite{Braun:2009, Erat:2009, Erat:2010, Erat:2010b}, finding an exponential relationship between conductivity and the relative spectral weight for the hole states versus that of the hybridized Fe($3d$)--O($2p$) states in the valence bands.

It is generally accepted that the heterovalent substitution of the $A$-site, namely La$^{3+}$ by Sr$^{2+}$, changes the $3d$ electronic configuration of TMs (in this case, a change in oxidation of Fe from 3+ to 4+), thereby providing a substitution parameter to control the valency of TM ions~\cite{Zhou:2005}. Moreover, such a substitution leads to the formation of hole states above the valence band, which has a substantial effect on the electronic transport properties. Due to the formation of doped hole states, the lowest energy charge excitation in the parent insulating oxides of the late transition metals is thus proven to be of the charge-transfer type~\cite{Chainani:1993a}. Information on the electronic structure of La$_{1-x}$Sr$_x$FeO$_3$ (LSFO) has been obtained using x-ray absorption near edge structure (XANES) spectroscopy~\cite{Abbate:1992}, which is sensitive to the local environment of the absorbing atom~\cite{Fuggle:1991, Tanaka:2005}: the Fe($2p$) spectrum of LFO is consistent with a $3d^5$ ($^6A_{1g}$) ground state, with a second component appearing in the Fe($2p$) spectra at higher Sr concentrations. Furthermore, the oxygen $K$-edge spectra of LSFO show a strong pre-peak appearing below the bottom of the conduction band of LFO due to the hole doping. Thus, Sr substitutions create new empty states (the so called ``hole doped peak'') above the top of the valence bands~\cite{Erat:2010}. This finding was confirmed by Chainani~{\it et al.}~\cite{Chainani:1993}, who also found that substitution of Sr$^{2+}$ in place of La$^{3+}$ introduce holes in the system and that the doped hole states have a mixed Fe($3d$)--O($2p$) character. Wadati~{\it et al.}~\cite{Wadati:2005} investigated the composition-dependent electronic structure of La$_{0.6}$Sr$_{0.4}$FeO$_3$ epitaxial thin films by {\it in-situ} photoemission spectroscopy (PES) and XANES measurements, and their band structure using angle-resolved photoemission spectroscopy (ARPES); tight-binding band-structure calculations with an empirical Hubbard $U$ could reproduce the overall behavior upon substitution, although they failed in explaining the hole-induced states above the Fermi energy~\cite{Wadati:2006}. The mechanism of $A$-site substitutions is, thus, relatively well understood.

Another class of compounds in this family involves homovalent substitutions at the $B$ site~\cite{Sarma:1994, Sarma:1998, Chainani:1996, Idrees:2011}.
% Jana:2019, Phuyal:2020
In contrast to the aforementioned $A$-site heterovalent substitution, the effects on valence and conduction bands deriving from the B-site homovalent substitution are less understood. For example, Sarma {\it et al.}~\cite{Sarma:1994} investigated homovalent B-site substitutions (with Mn, Fe and Co) for LaNiO$_3$ using XANES, and this study evidenced that B-site substitutions lead to redistribution of empty states without creation of new ones, contrary to what happens in the case of A-site substitutions. However, no specific explanation about the atomistic origin of this empty state redistribution was given. More recently, Idrees~{\it et al.}~\cite{Idrees:2011} investigated LaFe$_{1-x}$Ni$_x$O$_3$ ($0 \le x \le 0.5$) using XANES and impedance spectroscopic techniques, and they concluded that the lowest-energy empty states are of the Ni($3d$)--O($2p$) hybridized character, but no theoretical confirmation of such an analysis was given. Moreover, the B-site homovalent substitution could lead also to redistribution of occupied states, which could be probed by photoemission or x-ray emission spectroscopies. Therefore, theoretical and computational studies of LaFe$_{1-x}$Ni$_x$O$_3$ are highly desired in order to shed more light on our understanding of its electronic structure, XANES spectra, and thus to understand better the redistribution of empty and occupied states due to substitution of Ni for Fe. 
 
First-principles simulations of the ground-state structural, magnetic, and electronic properties of TM compounds (and ABO$_3$ perovskites in particular) are challenging due to the simultaneous presence of itinerant and localized electrons~\cite{Martin:2016}. Density-functional theory (DFT)~\cite{Hohenberg:1964, Kohn:1965} with its standard local-density and generalized-gradient approximations (LDA and GGA, respectively) to the exchange-correlation functional is by far the most widely used approach in condensed-matter physics and materials science for simulations of a vast variety of materials' properties. Nevertheless, and notwithstanding its numerous successes, it often fails to provide accurate description of TM oxides, not only quantitatively but often even qualitatively (e.g., it predicts a metallic instead of an insulating ground state in some TM oxides~\cite{Mandal:2019}). The failure of ``standard DFT'' is related first and foremost to very large self-interaction errors~\cite{Perdew:1981, MoriSanchez:2006} for localized $d$ and $f$ electrons~\cite{Kulik:2006, Kulik:2008, Kulik:2011}. Various approaches exist to alleviate these errors from DFT using (extended) Hubbard functionals (DFT+$U$ and DFT+$U$+$V$)~\cite{Anisimov:1991, Anisimov:1997, Dudarev:1998, Campo:2010, Himmetoglu:2014}, hybrid functionals~\cite{Stephens:1994, Heyd:2003}, or dynamical mean-field theory (DFT+DMFT)~\cite{Georges:1996, Anisimov:1997b, Lichtenstein:1998, Kotliar:2006} which also addresses this in its static limit. Each of these methods presents advantages and disadvantages, but we believe that for many materials' properties DFT+$U$ and DFT+$U$+$V$ provide the best compromise between computational cost and accuracy. The main rationale behind these methods is that the Hubbard term corrects selectively self-interaction errors in localized electrons using projections on the corresponding atomic manifolds, while itinerant electrons ($s$ and $p$) are treated at the LDA or GGA level. However, the values of the Hubbard parameters, which determine the strength of the correction, are unknown \textit{a priori}. While empirical evaluations of $U$ are quite common, various methods exist to compute it from first-principles~\cite{Springer:1998, Aryasetiawan:2004, Cococcioni:2005, Timrov:2018}, which render this method fully {\it ab initio}. In particular, an extension to include the effects of the inter-site $V$ electronic interactions (DFT+$U$+$V$) has been introduced quite recently~\cite{Campo:2010}, with the inter-site term being crucial for the energetics of complex TM oxides with covalent interactions (i.e., with significant hybridization between, e.g., the $d$ and $p$ states of neighboring sites)~\cite{Cococcioni:2019, Ricca:2020}. 

When a ground state of complex TM oxides is accurately described using, e.g., one of the aforementioned techniques, it is useful to perform spectroscopic investigations in order to further gain valuable information on various properties. In particular, the electronic structure can be investigated using x-ray absorption spectroscopy which involves the excitation of a core electron to the conduction band, leaving a hole in the core level. For delocalized edges, like $K$ and $L_1$, the description of the excitation requires calculating the density of empty states for a wide range of energies and, generally, in the presence of a core hole. Despite the fact that many techniques have been developed for modeling core-hole spectroscopy -- such as crystal-field multiplet theory~\cite{Cowan:1981}, real-space multiple scattering theory~\cite{Natoli:1980, Ankudinov:1998, Rehr:2010, Ebert:1996}, real-space finite differences~\cite{Joly:2001}, the DFT-based reciprocal-space approach~\cite{Taillefumier:2002,Gougoussis:2009b}, the DFT-based multiresolution approach~\cite{Ratcliff:2019}, and the approach based on solution of the Bethe-Salpeter equation of many-body perturbation theory~\cite{Shirley:1998, Soininen:2001} -- not all of these are equally applicable to the case of $K$ and $L_1$ edges, where the requirement of describing a very large set of states in a broad energy range eliminates the possibility of using either an effective Hamiltonian approach (such as crystal-field multiplet theory) or approaches that are computationally not affordable. 
DFT-based techniques are then the method of choice for the description of $K$ and $L_1$ edges; however, as was discussed before, due to the large self-interactions errors in standard functionals, conventional approaches for TM oxides involve the use of DFT+$U$~\cite{Cococcioni:2005}; this has provided in the past excellent results for the Ni $K$-edge XANES spectra in NiO~\cite{Gougoussis:2009}. Nevertheless, in TM oxides with strong covalent interactions, DFT+$U$ may not be sufficient and generalizations to include inter-site Hubbard interactions could be considered.

In this work we present a joint theoretical and experimental study of the oxygen $K$-edge spectra of LFO and LaFe$_{0.75}$Ni$_{0.25}$O$_{3}$ (LFNO); this corresponds to transitions from the O $1s$ core states to O unoccupied $p$ states. By extending the XANES approach of Ref.~\cite{Gougoussis:2009} to a Hubbard-corrected approach that includes both on-site (Hubbard $U$) and inter-site (Hubbard $V$) electronic interactions we explore the effects of a more explicit and accurate account of these electronic interactions
(and of the consequent alleviation of the electronic self-interactions) on the XANES spectra, whose quality is assessed by direct comparison with spectroscopic data. The present implementation of DFT+$U$+$V$ allows for calculations of XANES spectra in a broad energy range thanks to the use of the  Lanczos recursive method, which removes the need of computing empty states which otherwise would be needed when standard DFT-based techniques are used. As it will be shown, the use of extended Hubbard functionals allows us to refine some low-energy features in the XANES spectra at a computational cost comparable to standard DFT~\cite{TancogneDejean:2019, Ricca:2020}, finding that the pre-peak in the oxygen $K$-edge spectrum of LFNO is of mixed Ni($3d$)--O($2p$) character, in agreement with experimental findings of Refs.~\cite{Erat:2010, Idrees:2011}. Other parts of the spectra are instead left substantially unchanged with respect to DFT+$U$.

The paper is organized as follows: In Sec.~\ref{sec:methods} we present a description of the computational approach used and details of the simulations and measurements; in Sec.~\ref{sec:PDOS} we discuss the projected density of unoccupied states; in Sec.~\ref{sec:XANES} we present a comparison of the calculated and measured oxygen $K$-edge spectra of LFO and LFNO, and finally in Sec.~\ref{sec:Conclusions} we provide our conclusions. The Appendix~\ref{secSM:PDOS_valence} contains additional information describing the projected density of occupied states, and Appendix~\ref{secSM:noUonLa4f} presents a discussion of the role of the Hubbard $U$ correction for the La($4f$) states.

\section{Methods}
\label{sec:methods}

\subsection{Computational approach}
\label{sec:comput_approach}

In this section we discuss the computational approach which we develop and use to compute the XANES spectra of LFO and LFNO using extended Hubbard functionals. For the sake of simplicity, we present the equations for norm-conserving pseudopotentials, but the generalization to ultrasoft pseudopotentials is straightforward and would follow Refs.~\cite{Gougoussis:2009, Timrov:2020}.

According to Fermi's golden rule, the x-ray absorption cross section is defined as:
\begin{equation}
    \sigma(\omega) = 4 \pi^2 \alpha_0 \hbar \omega \sum_{f,\mathbf{k},\sigma} |\langle \psi^\sigma_{f,\mathbf{k}}| \mathfrak{D} | \psi^\sigma_{i,\mathbf{k}} \rangle|^2 \delta(\varepsilon^\sigma_{f,\mathbf{k}} - \varepsilon^\sigma_{i,\mathbf{k}} - \hbar\omega) \,,
    \label{eq:cross_section}
\end{equation}
where $\alpha_0$ is the fine-structure constant, $\hbar\omega$ is the photon energy, $\mathbf{k}$ are the points in the Brillouin zone, $\sigma$ is the spin index, $\psi^\sigma_{i,\mathbf{k}}$ and $\psi^\sigma_{f,\mathbf{k}}$ are the wavefunctions of the initial $i$ and final $f$ states, and $\varepsilon^\sigma_{i,\mathbf{k}}$ and $\varepsilon^\sigma_{f,\mathbf{k}}$ are their energies. The transition amplitudes between the initial and final states are defined as matrix elements $\langle \psi^\sigma_{f,\mathbf{k}} | \mathfrak{D} | \psi^\sigma_{i,\mathbf{k}} \rangle$, where $\mathfrak{D} = \mathbf{e} \cdot \mathbf{r}$ is the transition operator in the dipole approximation with $\mathbf{e}$ and $\mathbf{r}$ being the polarization vector of the photon beam and the electron position vector.

The cross section given by Eq.~\eqref{eq:cross_section} can be evaluated efficiently as a continued fraction using a recursive Lanczos method, avoiding the explicit calculation of unoccupied states (as explained in Refs.~\cite{Taillefumier:2002, Gougoussis:2009b}).

In the case of TM oxides, the self-interaction contribution present in DFT with LDA or GGA functionals tends to overdelocalize $d$ and $f$ electrons, with negative consequences on the ground-state structural and electronic properties, and thus on the computed XANES spectra. DFT+$U$ and DFT+$U$+$V$ aim to address this problem~\cite{Anisimov:1991, Dudarev:1998, Kulik:2006, Kulik:2008, Campo:2010, Kulik:2011, Himmetoglu:2014}. 
The main idea is to define the total energy as
\begin{equation}
E_{\mathrm{DFT}+U+V} = E_{\mathrm{DFT}} + E_{U+V} \,,
\label{eq:Edft_plus_u}
\end{equation}
where $E_{\mathrm{DFT}}$ is the approximate DFT energy [constructed, e.g., within the local spin density approximation (LSDA) or the spin-polarized generalized gradient approximation ($\sigma$-GGA)], while $E_{U+V}$ contains the difference between the Hubbard term and its mean-field approximation, subtracted to avoid the double-counting of interactions already included in $E_{\mathrm{DFT}}$. In the present work, this latter term is shaped according to the popular fully localized limit~\cite{Anisimov:1997, Dudarev:1998, Himmetoglu:2014}. At variance with the simpler DFT+$U$ approach, 
containing only on-site interactions, DFT+$U$+$V$ is based on the extended Hubbard model including also inter-site interactions:
\begin{eqnarray}
E_{U+V} & = & \frac{1}{2} \sum_I \sum_{\sigma m m'} 
U^I \left( \delta_{m m'} - n^{II \sigma}_{m m'} \right) n^{II \sigma}_{m' m} \nonumber \\
& & - \frac{1}{2} \sum_{I} \sum_{J (J \ne I)}^* \sum_{\sigma m m'} V^{I J} 
n^{I J \sigma}_{m m'} n^{J I \sigma}_{m' m} \,,
\label{eq:Edftu}
\end{eqnarray}
where $I$ and $J$ are the atomic site indices, $m$ and $m'$ are the magnetic quantum numbers associated with a specific angular momentum, $U^I$ and $V^{I J}$ are the on-site and inter-site Hubbard parameters
($U^I \equiv V^{I I}$), and the star in the second sum in Eq.~\eqref{eq:Edftu} means that for each atom $I$
the index $J$ covers all its nearest neighbors up to a given distance.

By taking a functional derivative of $E_{\mathrm{DFT}+U+V}$ with respect to the complex conjugate of Kohn-Sham wavefunctions, the Kohn-Sham equations including the Hubbard corrections are obtained:
\begin{equation}
    \hat{H}^\sigma \psi^\sigma_{v,\mathbf{k}} = \varepsilon^\sigma_{v,\mathbf{k}} \psi^\sigma_{v,\mathbf{k}} \,,
\end{equation}
where $v$ is the electronic band index, and $\hat{H}^\sigma$ is the total Hamiltonian of the unperturbed system. For the generalized DFT+$U$+$V$ functional one has
\begin{equation}
    \hat{H}^\sigma = \hat{H}^\sigma_\mathrm{DFT} + \hat{V}^\sigma_U + \hat{V}^\sigma_V \,,
    \label{eq:Hamiltonian}
\end{equation}
where $\hat{H}^\sigma_\mathrm{DFT}$ is the DFT Hamiltonian, $\hat{V}^\sigma_U$ is the Hubbard potential associated with the on-site correction
\begin{equation}
    \hat{V}^\sigma_U = \sum_I \sum_{m m'} U^{I} \left( \frac{\delta_{m m'}}{2} - 
n^{I \sigma}_{m m'} \right) |\varphi^I_{m} \rangle \langle \varphi^I_{m'}| \,,
    \label{eq:potU}
\end{equation}
and $\hat{V}^\sigma_V$ is the Hubbard potential associated with the inter-site interactions between neighboring atoms
\begin{equation}
    \hat{V}^\sigma_V = - \sum_{I} \sum_{J (J \ne I)}^* \sum_{m m'} V^{I J} 
    n^{I J \sigma}_{m m'} |\varphi^I_{m} \rangle \langle \varphi^J_{m'}| \,,
    \label{eq:potV}
\end{equation}
In Eqs.~\eqref{eq:potU} and \eqref{eq:potV}, $\varphi^I_{m}$ and $\varphi^J_{m'}$ are the localized orbitals, and $n^{I J \sigma}_{m m'}$ are the generalized atomic occupation matrices
\begin{equation}
   n^{I J \sigma}_{m m'} = \sum_{v,\mathbf{k}} \, f^\sigma_{v,\mathbf{k}} \langle \psi^\sigma_{v,\mathbf{k}} | \varphi^J_{m'} \rangle
   \langle \varphi^I_{m} | \psi^\sigma_{v,\mathbf{k}}\rangle \,, 
   \label{eq:occ_matrix}
\end{equation}
where $f^\sigma_{v,\mathbf{k}}$ are the occupations of Kohn-Sham states. In Eq.~\eqref{eq:potU}, $n^{I \sigma}_{m m'} \equiv n^{I I \sigma}_{m m'}$~\cite{Note:equations}. 

In DFT+$U$ (and DFT+$U$+$V$) the values of the Hubbard parameters are not known {\it a priori}, and often their values are adjusted semiempirically, by matching the value of properties of interest. This is fairly arbitrary, and in addition it is often based on properties, such as the band gap, that are outside the scope of even exact DFT. In this work we instead determine Hubbard parameters from a generalized piece-wise linearity condition implemented through linear-response theory~\cite{Cococcioni:2005}, based on density-functional perturbation theory (DFPT)~\cite{Timrov:2018, Timrov:2020}. Within this framework the Hubbard parameters are the elements of an effective interaction matrix computed as the difference between bare and screened inverse susceptibilities~\cite{Cococcioni:2005}:
\begin{equation}
U^I = \left(\chi_0^{-1} - \chi^{-1}\right)_{II} \,,
\label{eq:Ucalc}
\end{equation}
\begin{equation}
V^{IJ} = \left(\chi_0^{-1} - \chi^{-1}\right)_{IJ} \,,
\label{eq:Vcalc}
\end{equation}
where $\chi_0$ and $\chi$ are the susceptibilities which measure the response of atomic occupations to shifts in the potential acting on individual Hubbard manifolds. In particular, $\chi$ is defined as $\chi_{IJ} = \sum_{m\sigma} \left(dn^{I \sigma}_{mm} / d\alpha^J\right)$, where $\alpha^J$ is the strength of the perturbation on the $J^\mathrm{th}$ site. While $\chi$ is evaluated at self-consistency (of the linear-response Kohn-Sham calculation), $\chi_0$ (which has a similar definition) is computed before the self-consistent re-adjustment of the Hartree and exchange-correlation potentials~\cite{Cococcioni:2005}. The main goal of the DFPT implementation is to recast the response to such isolated perturbations in supercells as a sum over a regular grid of $N_\mathbf{q}$ ${\mathbf{q}}$ points in the Brillouin zone~\cite{Timrov:2018}
\begin{equation}
\frac{dn^{I \sigma}_{mm'}}{d\alpha^J} = \frac{1}{N_{\mathbf{q}}}\sum_{\mathbf{q}}^{N_{\mathbf{q}}} e^{i\mathbf{q}\cdot(\mathbf{R}_{l} - \mathbf{R}_{l'})}\Delta_{\mathbf{q}}^{s'} n^{s \sigma}_{mm'} \,,
\label{eq:dnq}
\end{equation}
in order to exploit the efficiency of DFPT (the monochromatic responses in Eq.~\eqref{eq:dnq} are calculated in the primitive cell). In Eq.~\eqref{eq:dnq}, the atomic indices have been replaced by atomic ($s$ and $s'$) and unit cell ($l$ and $l'$) labels [$I\equiv(l,s)$ and $J\equiv(l',s')$]. $\mathbf{R}_l$ and $\mathbf{R}_{l'}$ are the Bravais lattice vectors and the grid of $\mathbf{q}$ points is chosen fine enough to make the resulting atomic perturbations effectively decoupled from their periodic replicas~\cite{Timrov:2018}. Since the $\Delta_{\mathbf{q}}^{s'} n^{s \sigma}_{mm'}$ are the lattice-periodic responses of atomic occupations to a monochromatic perturbation of wavevector $\mathbf{q}$, they can be obtained by solving the DFPT equations independently for every $\mathbf{q}$~\cite{Timrov:2018, Timrov:2020}; as mentioned before, this allows us to eliminate the need of supercells when computing $U$ and $V$.

\subsection{Computational details}
\label{sec:comput_details}

All calculations are performed using the open-source \QE{} 
distribution~\cite{Giannozzi:2009, Giannozzi:2017, Giannozzi:2020}, whose extended DFT+$U$+$V$ implementation, employed in this work, is now publicly available to the community at large~\cite{QuantumESPRESSO:Gitlab} and has been officially released with version~6.6 of the distribution~\cite{QuantumESPRESSO:website}. Plane-wave basis sets and the pseudopotential approximation are used~\cite{Martin:2004}. We have used $\sigma$-GGA for the exchange-correlation functional constructed with the Perdew-Burke-Ernzerhof (PBE) parametrization~\cite{Perdew:1996}, and ultrasoft pseudopotentials from the Pslibrary~0.3.1 and 1.0.0~\cite{Dalcorso:2014, Note:Pseudopotentials}. Kohn-Sham wave functions and charge density are expanded in plane waves up to a kinetic-energy cutoff of 60 and 480~Ry, respectively; spin-orbit coupling is neglected.

\begin{figure}[t]
 \includegraphics[width=1.0\linewidth]{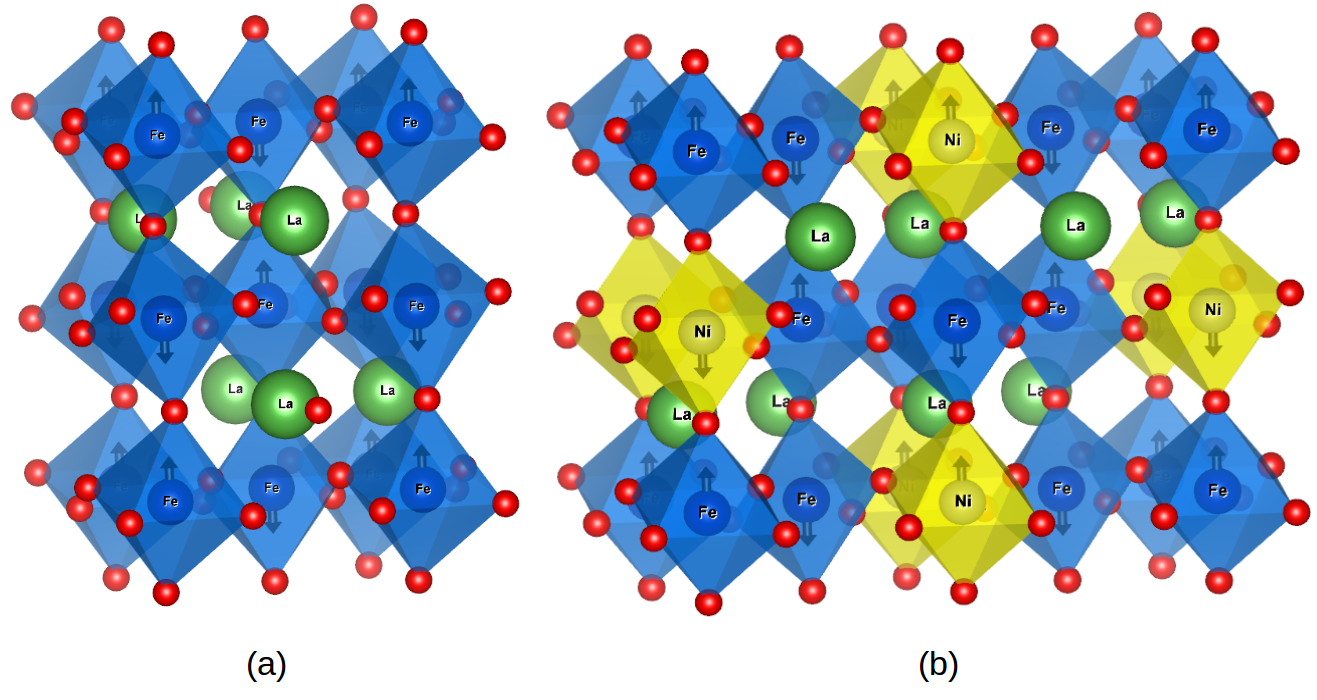}
\caption{Crystal structure of (a)~LaFeO$_3$, and (b)~LaFe$_{0.75}$Ni$_{0.25}$O$_3$.
Both structures are of the G-type AFM ordering (vertical arrows indicate the orientation
of the spin on each atom). Color code: La atoms (green), Fe atoms (blue), 
Ni atoms (yellow), O atoms (red), blue and yellow octahedra are centered on Fe and Ni 
atoms, respectively. Figures were produced using the VESTA program~\cite{Momma:2011}.}
\label{fig:cryst_structure}
\end{figure}

According to the Rietveld analysis~\cite{Larson:1994, Toby:2001} of the x-ray diffractograms, 
both LFO and LFNO have orthorhombic symmetry with space group $Pnma$ (No.~62)~\cite{Erat:2010b}. 
We have used the experimental lattice parameters and atomic positions for LFO and LFNO as measured in this work~\cite{MaterialsCloudData}. LFO and LFNO have the G-type AFM configuration below the N\'eel temperature, according to the experimental results~\cite{Hearne:1995, Falcon:1997} and calculations~\cite{Sarma:1995}; this is confirmed by our calculations, and therefore we used this magnetic ordering. The unit cells for ground-state calculations consist of 20 and 40 atoms for LFO and LFNO, respectively, and are shown in Fig.~\ref{fig:cryst_structure}. For LFNO this cell contains two substitutional Ni impurities that are placed at the maximum distance from one another.

\begin{table*}[t]
 \begin{center}
  \begin{tabular}{lccccccccc}
    \hline\hline
                  &                \multicolumn{2}{c}{LaFeO$_3$}                      & \phantom{a} &                                        \multicolumn{4}{c}{LaFe$_{0.75}$Ni$_{0.25}$O$_{3}$}                              \\ \cline{2-4} \cline{6-10}
                  &  $U_{\mathrm{Fe}(3d)}$ & $U_{\mathrm{La}(4f)}$ & $V_{\mathrm{Fe}(3d) - \mathrm{O}(2p)}$  &  &$U_{\mathrm{Fe}(3d)}$  &  $U_{\mathrm{La}(4f)}$ & $U_{\mathrm{Ni}(3d)}$      & $V_{\mathrm{Fe}(3d) - \mathrm{O}(2p)}$ & $V_{\mathrm{Ni}(3d) - \mathrm{O}(2p)}$   \\ \hline
DFT+$U$           &         5.16           &        3.40           &                 --                      &  &    5.28/5.36         &            3.41        &        7.23                &           --                           &               --                        \\  
DFT+$U$+$V$       &         5.54           &        3.22           &          0.77--0.78                     &  &    5.73/5.82         &            3.26        &        7.65                &         0.63--0.99                     &           0.67--0.98                    \\ \hline\hline
   \end{tabular}
 \end{center}
\caption{Self-consistent Hubbard parameters (in eV) for LaFeO$_3$ and LaFe$_{0.75}$Ni$_{0.25}$O$_{3}$ computed using DFPT~\cite{Timrov:2018, Timrov:2020} when using pseudopotentials from the Pslibrary~0.3.1 and 1.0.0~\cite{Dalcorso:2014, Note:Pseudopotentials}. In the case of LaFe$_{0.75}$Ni$_{0.25}$O$_{3}$ there are two inequivalent Fe sites, thus two values of $U$ for Fe($3d$) are specified. For both materials, the inter-site $V$ is specified as a range of values, because there are various inequivalent pairs of neighbors.}
\label{tab:HP}
\end{table*}

Calculations of the Hubbard parameters ($U$ and $V$) are performed using the \texttt{HP} code of \QE{} which is based on DFPT~\cite{Timrov:2018, Timrov:2020}. To construct the projectors on the Hubbard manifold we use orthogonalized atomic orbitals~\cite{Lowdin:1950, Mayer:2002}. The Hubbard interaction parameters are computed in a self-consistent way, i.e., through an iterative procedure that involves recomputing their values from the Hubbard-corrected ground state until they converge~\cite{Hsu:2009, Cococcioni:2019, Ricca:2019}. For LFO we used equal uniform $4 \times 4 \times 2$ $\mathbf{k}$ and $\mathbf{q}$ points meshes, and for LFNO we used $3 \times 1 \times 2$, which allowed us to converge Hubbard parameters $U$ and $V$ with the accuracy of about 0.01~eV. The computed values of Hubbard parameters for LFO and LFNO are listed in Table~\ref{tab:HP}. La sites have $4f$ and $5d$ unoccupied states: we computed Hubbard $U$ only for $4f$ states, because their energy is inaccurate in LDA/GGA, while we did not compute $U$ for $5d$ states (current implementation of the \texttt{HP} code does not allow us to use the Hubbard correction for more than one manifold of the same atom). Moreover, La($5d$) states are not very localized and hence the Hubbard $U$ correction is likely not so crucial~\cite{Nohara:2009}. As will be shown in Sec.~\ref{sec:results}, the use of the Hubbard $U$ only on La($4f$) states turns out to be sufficient to explain the origin of all the peaks in the XANES spectra of LFO and LFNO. The inter-site Hubbard $V$ between La($4f$) and O($2p$) was found to be very small ($<0.1$~eV); therefore, it was neglected in the calculations. The projected density of states (PDOS) is computed using uniform $\mathbf{k}$ points meshes ($12 \times 12 \times 8$ for LFO and $12 \times 6 \times 8$ for LFNO) and by summing over contributions from all sites; to plot the PDOS we used a Gaussian function with a broadening parameter of 0.02~Ry.

XANES calculations are performed using the \texttt{XSpectra} code of \QE{}, which is based on the evaluation of Eq.~\eqref{eq:cross_section} in reciprocal space on top of DFT (or DFT+$U$, or DFT+$U$+$V$) results, using the Lanczos recursive algorithm~\cite{Taillefumier:2002, Gougoussis:2009}. We used an approximation which consists of neglecting the core hole left behind by the O($1s$) electron promoted to the conduction manifold. This is known to be a good approximation for the O $K$-edge spectra of LaFeO$_3$~\cite{Wu:1997} (in addition, our preliminary tests have shown that the effect of the core hole is not crucial for the interpretation of our experimental XANES spectra). All XANES spectra are computed in the dipole approximation (i.e. by neglecting a quadrupole and higher-order terms). The XANES intensities are computed as a mean average over three perpendicular polarizations of the photon beam (i.e. along Cartesian axes $x$, $y$, $z$), which in turn are computed for all inequivalent O atoms. To converge O $K$-edge spectra we use the same uniform $\mathbf{k}$ points meshes of the PDOS.
The XANES spectral lines are convoluted with a Lorentzian smearing function having an energy-dependent broadening parameter~\cite{Bunau:2013}, which starts from 0.16~eV (the linewidth for O($1s$)~\cite{Sankari:2003}) and reaches a maximum value of 2.00~eV, with an arctan-type behavior and inflection point at 12~eV above the top of the valence-band maximum~\cite{Bunau:2013}. 

\subsection{Experimental details}
\label{sec:exp_details}

LFO and LFNO were prepared by conventional solid-state reactions. The precursors 
La$_2$O$_3$ ($>99.99$\%), Fe$_2$O$_3$ ($>99.0$\%), and NiO ($99.8$\%) were mixed in stoichiometric 
proportions, calcined at 1200~$^\circ$C for 4~h and then sintered at 1400~$^\circ$C for 12~h
with a heating and cooling rate of 5~K/min for LFNO, and LFO was calcined at 1200~$^\circ$C 
for (4~h + 4~h) with the same heating and cooling rate. 
The x-ray powder diffractograms of LFO and LFNO were collected with a Philips X'Pert PRO-MPD diffractometer at ambient temperature (40 kV, 40 mA, Cu K$_\alpha$ $\lambda=1.5405$~\AA) in steps of 0.02$^\circ$ for $20^\circ \leq 2 \Theta  \leq 80^\circ$. The oxygen deficiency $\delta$ of the samples was obtained by thermogravimetry~\cite{Erat:2009}.

XANES spectra at 300~K were recorded at the Advanced Light Source in Berkeley, beamline~9.3.2, the end station of which has an operating energy range 200--1200~eV and an energy resolution of 1/10000. The vacuum chamber base pressure was lower than $5 \times 10^{-10}$~Torr. Signal detection was made in total electron yield mode. Powder samples were dispersed on conducting carbon tape and then mounted on a copper sample holder. Oxygen $K$-edge spectra were recorded from 520 to 560~eV in steps of 0.1~eV. An arctan function, which accounts for the absorption above the ionization threshold, was subtracted from the experimental oxygen $K$-edge spectrum; this allows for a more facile comparison of experimental spectra with calculated ones.

\section{Results and discussion}
\label{sec:results}

\subsection{Electronic structure}
\label{sec:PDOS}

The PDOS is the simplest approximation and a very useful tool to interpret XANES spectra. Since XANES probes empty states, and since the main goal of this study is the interpretation of the XANES spectra of LFO and LFNO, we will discuss only the projected density of empty states, with the main focus on O($2p$) states (see Appendix~\ref{secSM:PDOS_valence} for the PDOS for occupied states). 

\begin{table}[h]
 \begin{center}
  \begin{tabular}{lccc}
    \hline\hline
                  &    \parbox{2cm}{LaFeO$_3$}     &  \multicolumn{2}{c}{\parbox{3cm}{LaFe$_{0.75}$Ni$_{0.25}$O$_{3}$}}  \\ \hline
                  & \parbox{1cm}{$|m|_\mathrm{Fe}$} & \parbox{1cm}{$|m|_\mathrm{Fe}$} & \parbox{1cm}{$|m|_\mathrm{Ni}$}  \\ \hline
     DFT                 &        3.77                            &      3.53             &      0.30          \\       
     DFT+$U$             &        4.14                            &      4.08             &      1.87          \\
     DFT+$U$+$V$         &        4.10                            &      4.05             &      1.86          \\
     Expt.               &     3.9~\cite{Zhou:2006}               &      N/A              &      N/A           \\
                         &     $4.6 \pm 0.2$~\cite{Koehler:1957}  &                       &                    \\
     \hline\hline
  \end{tabular}    
 \end{center}
\caption{Average total magnetic moments per element (in $\mu_\mathrm{B}$) obtained using DFT, DFT+$U$, and DFT+$U$+$V$, and as measured in experiments~\cite{Zhou:2006,Koehler:1957}. ``N/A'' stands for ``not available''.}
\label{tab:magnetic_moments}
\end{table}

In Table ~\ref{tab:magnetic_moments} we report the magnetic moments in LFO and LFNO computed at different levels of theory, and compare them with the available experimental data. It can be seen that in the case of LFO the inclusion of the Hubbard $U$ (and $V$) corrections increases the value of the magnetic moment for Fe by about 10\% with respect to standard DFT, and thus the overall agreement with the experimental magnetic moments of Refs.~\cite{Zhou:2006,Koehler:1957} is improved. It is instructive to make a comparison with DFT+$U$ studies of Ref.~\cite{Nohara:2009} which were performed by applying $U$ only to La($4f$) states and no $U$ correction for Fe$(3d)$ states. In Ref.~\cite{Nohara:2009} it was found that DFT+$U$ gives the magnetic moment of 3.54~$\mu_\mathrm{B}$, which is in looser agreement with experiments than our DFT+$U$ and DFT+$U$+$V$ results obtained with $U$ also on Fe($3d$) states (see Table~\ref{tab:HP}). This observation is not surprising, because neglecting $U$ on Fe($3d$) states does not make a very big difference in the magnetization of the system with the standard DFT, and application of the $U$ correction only to La($4f$) states does not influence noticeably the magnetic moments on Fe atoms. Furthermore, the more advanced approach of Ref.~\cite{Nohara:2009}, based on $GW$ calculations~\cite{Hedin:1965} performed on top of DFT+$U$, results in a magnetic moment of 3.37~$\mu_\mathrm{B}$ which  further worsens the agreement with the experimental values~\cite{Zhou:2006,Koehler:1957}. Therefore, the application of the Hubbard $U$ correction to Fe($3d$) states is important for the accurate description of magnetic moments in LFO. In the case of LFNO, Hubbard $U$ (and $V$) corrections increase the value of magnetic moment for Fe by about 15\% with respect to standard DFT. However, the magnetic moment for Ni in LFNO is increased by a factor of $\approx 6$ due to the inclusion of Hubbard corrections, which highlights the crucial role of the Hubbard $U$ correction for Ni($3d$) states. To the best of our knowledge, there are no experimental data for magnetic moments in LFNO, therefore we cannot validate the accuracy of our DFT+$U$ and DFT+$U$+$V$ magnetic moments in this material. Last, the differences between DFT+$U$ and DFT+$U$+$V$ results for magnetic moments in LFO and LFNO are very small, meaning that the inclusion of inter-site Hubbard interactions does not change significantly the magnetic moments with respect to the DFT+$U$ case. 

The band gaps of LFO and LFNO are reported in Table~\ref{tab:Egap}. A detailed discussion about the band gaps for each system will be given in the following. But we want to stress at this point that the goal of this work is not to obtain very accurate band gaps~\cite{Note:DFTgap}, but to interpret XANES experiments based on the analysis of the PDOS and simulated XANES spectra. 

\begin{table}[h]
 \begin{center}
  \begin{tabular}{lccccccc}
    \hline\hline
                                     & DFT   & & DFT+$U$  & & DFT+$U$+$V$ & & Expt. \\ \hline
     LaFeO$_{3}$                     & 0.65  & &  2.69    & &     2.96    & &  2.1  \\
     LaFe$_{0.75}$Ni$_{0.25}$O$_{3}$ & 0.00  & &  0.56    & &     1.00    & &  N/A  \\
     \hline\hline
  \end{tabular}    
 \end{center}
\caption{Energy band gaps (in eV) computed for LaFeO$_{3}$ and LaFe$_{0.75}$Ni$_{0.25}$O$_{3}$ using DFT, DFT+$U$, and DFT+$U$+$V$, and compared with the experimental value from Ref.~\cite{Arima:1993}. These band gaps cannot be easily seen in Fig.~\ref{fig:PDOS} due to the large broadening parameter used to smooth the PDOS (for the sake of easier comparison with the experimental XANES spectra). ``N/A'' stands for ``not available''.}
\label{tab:Egap}
\end{table}

\subsubsection{LaFeO$_3$}
\label{sec:PDOS_LFO}

\begin{figure*}[t]
\begin{center}
 \subfigure[]{\includegraphics[width=0.48\linewidth]{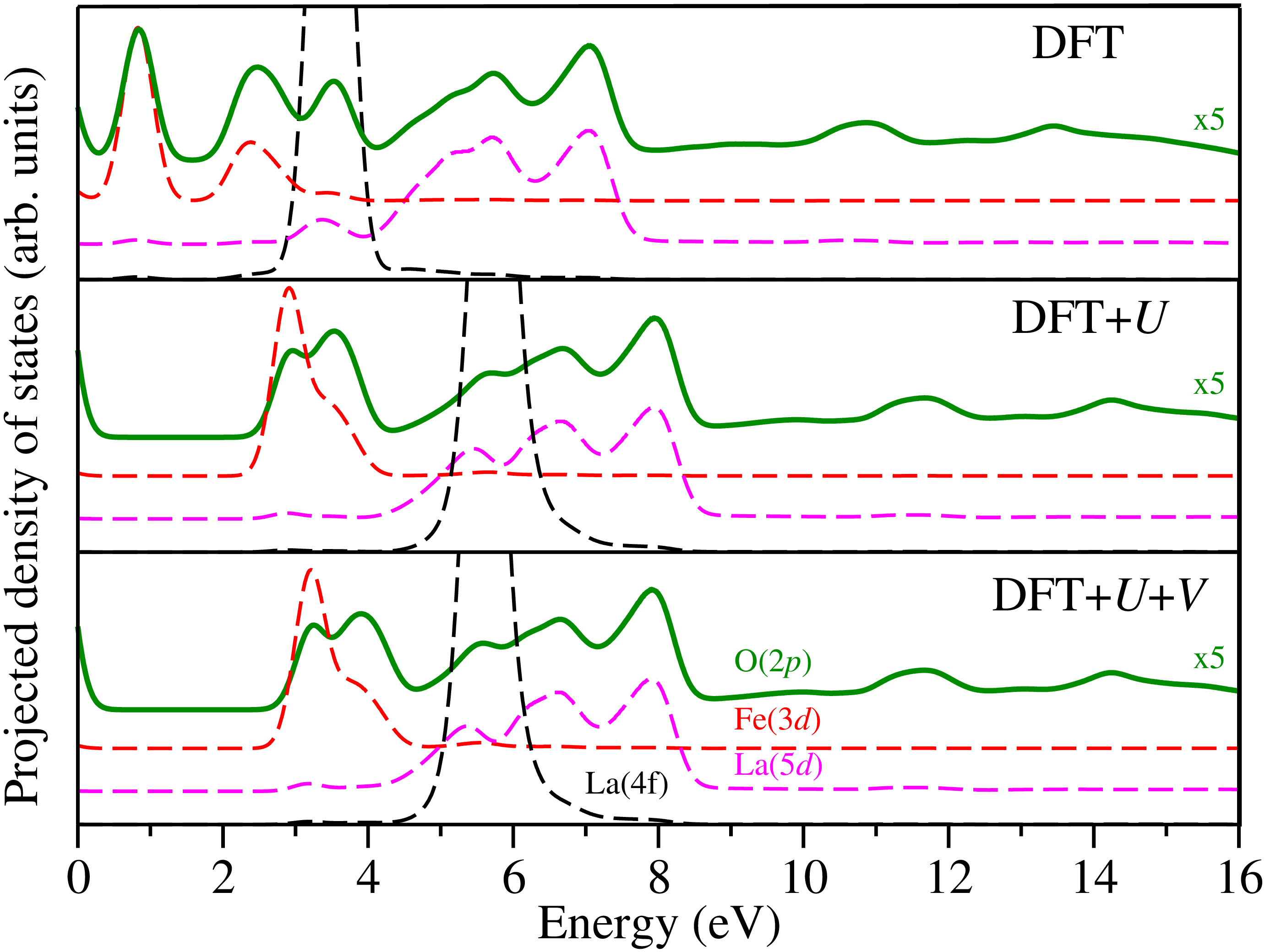}\hspace{0.05cm}}
 \subfigure[]{\includegraphics[width=0.48\linewidth]{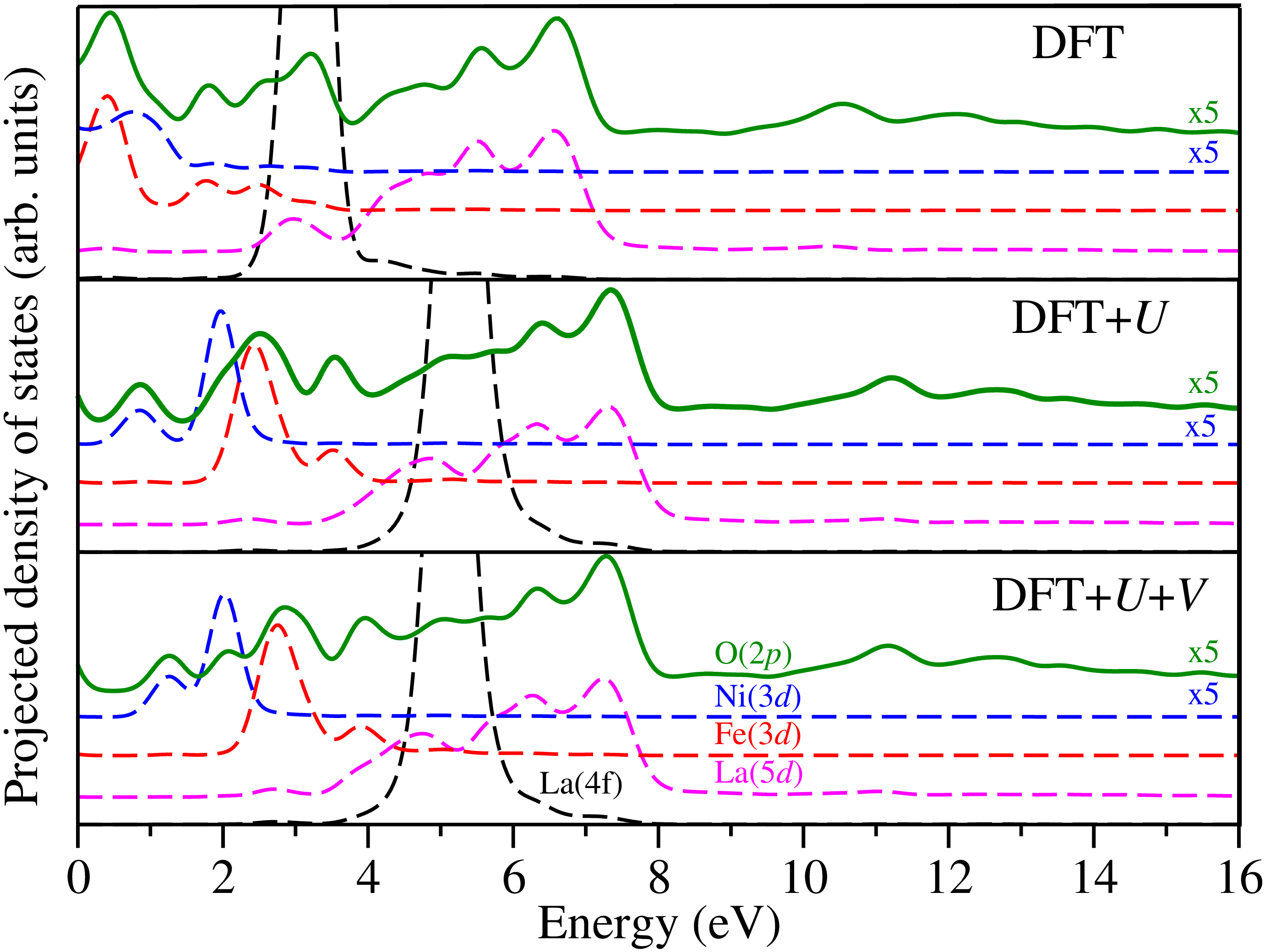}}
 \subfigure[]{\includegraphics[width=0.48\textwidth]{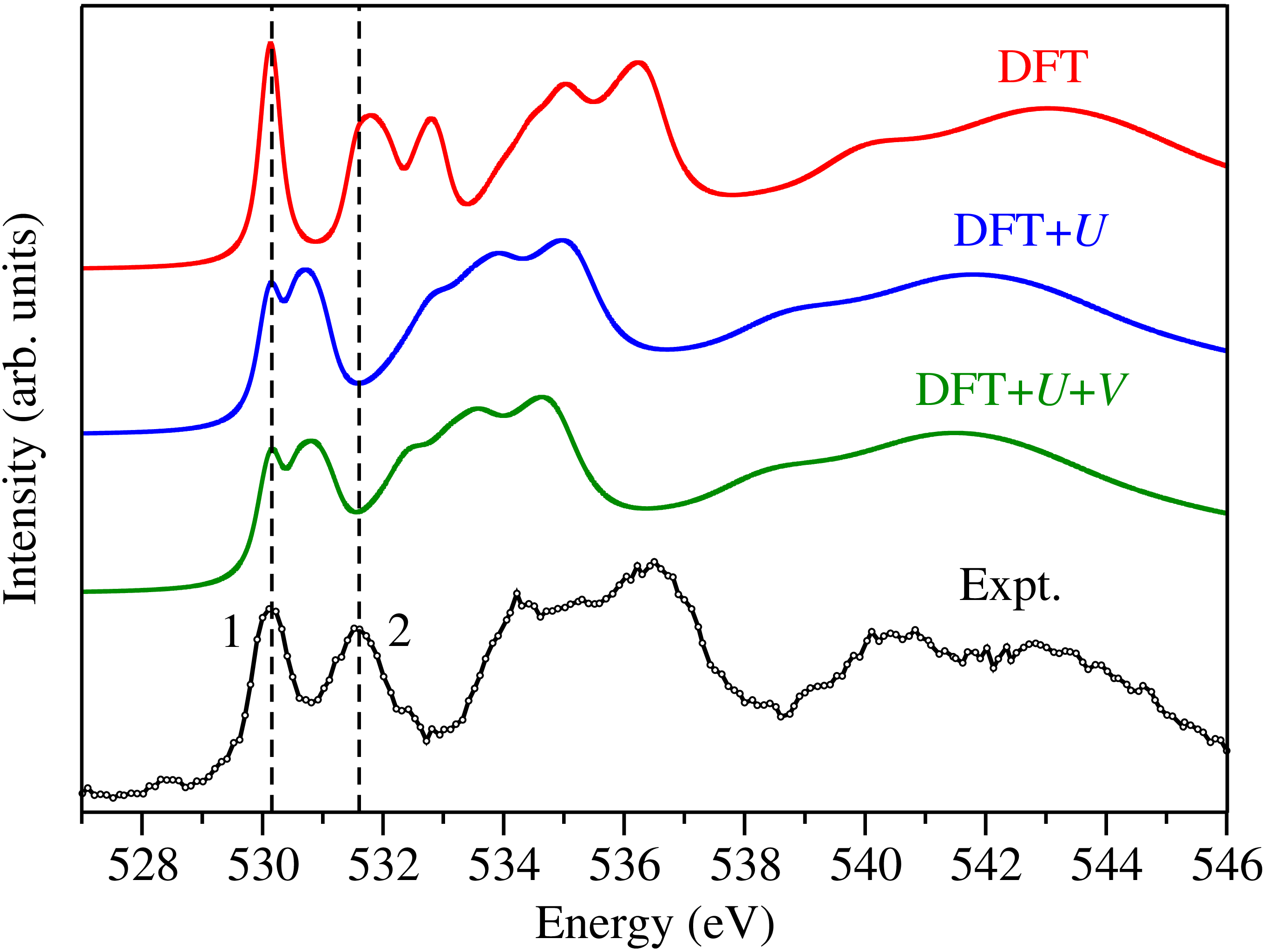}\hspace{0.05cm}}
 \subfigure[]{\includegraphics[width=0.48\textwidth]{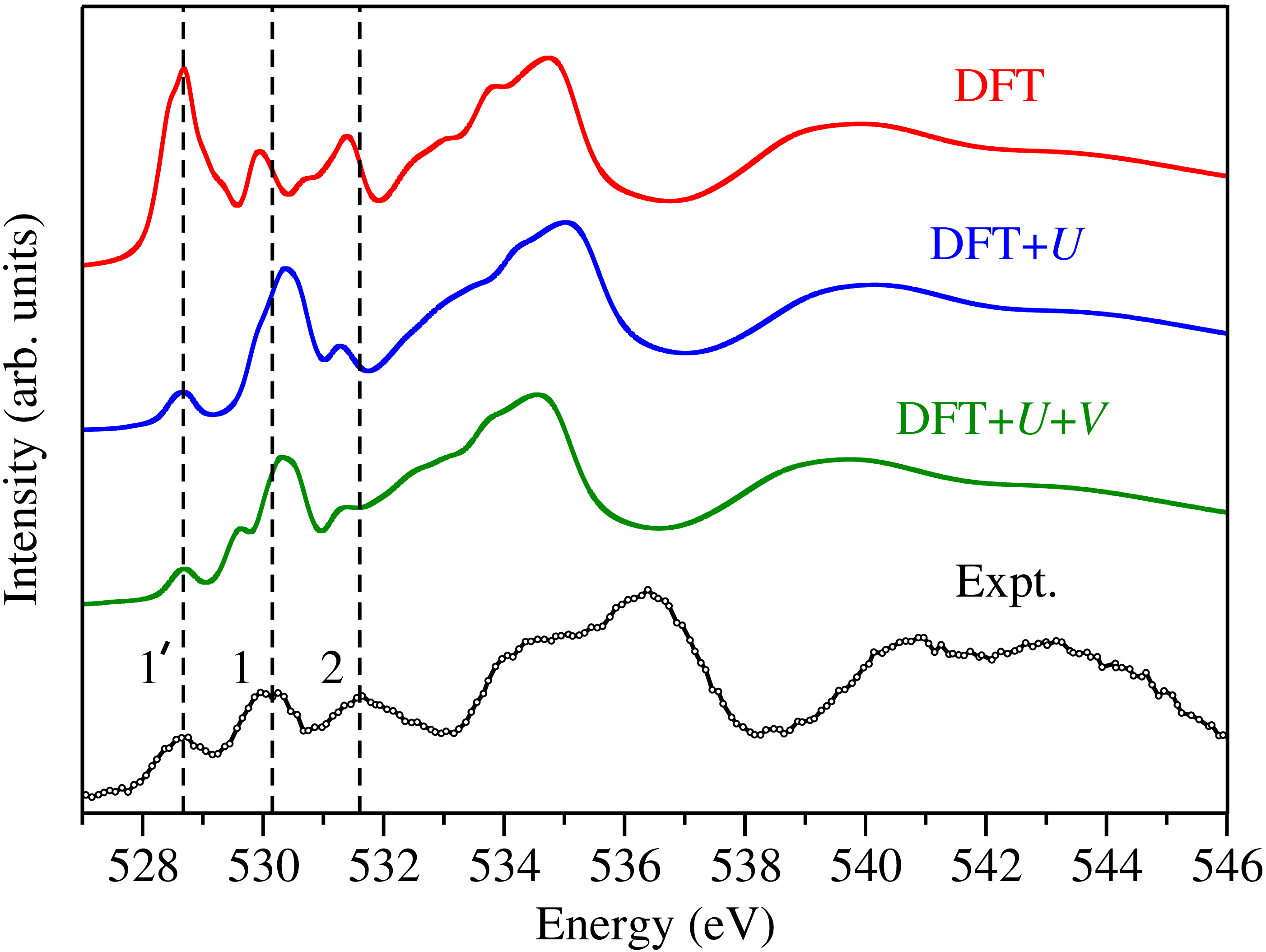}}
 \caption{Projected density of empty states computed using DFT, DFT+$U$, and
DFT+$U$+$V$ for (a)~LaFeO$_3$, and (b)~LaFe$_{0.75}$Ni$_{0.25}$O$_{3}$, using Hubbard parameters listed in Table~\ref{tab:HP}. The zero of energy corresponds to the valence band maximum (or Fermi energy in the case of LaFe$_{0.75}$Ni$_{0.25}$O$_{3}$ at the DFT level of theory, which comes out to be metallic). The PDOSs for O($2p$) and Ni($3d$) states are multiplied by a factor of 5 for easier comparison with other PDOS components; the PDOSs for Fe($3d$) and Ni($3d$) states show only the minority spin components on all panels since the majority spin components correspond to occupied states (see Appendix~\ref{secSM:PDOS_valence}). Comparison between the experimental and theoretical oxygen $K$-edge XANES spectra of (c) LaFeO$_3$, and (d) LaFe$_{0.75}$Ni$_{0.25}$O$_3$. The energy scales for the experimental spectra of LFO and LFNO are aligned (by a rigid shift) in order to facilitate the comparison. PDOS and XANES spectra are shifted vertically for clarity.}
\label{fig:PDOS}
\end{center}
\end{figure*}

In the case of LFO at the DFT level [see Fig.~\ref{fig:PDOS}~(a), top panel], the band gap is underestimated roughly by a factor of 3 with respect to the experimental value (see Table~\ref{tab:Egap}). As can be seen in Fig.~\ref{fig:PDOS}~(a) (top panel), the first and second peaks in the PDOS for O($2p$) appear at 0.8 and 2.5~eV, respectively, and they originate from hybridization with Fe($3d$) minority spin states. The two peaks for Fe($3d$) minority spin states (which are $t_{2g}$ and $e_g$) are due to the crystal-field splitting and the exchange splitting (the latter being much larger than the former)~\cite{deGroot:1989}, and hence the O($2p$) PDOS also reflects these two peaks due to the hybridization effects, in agreement with previous calculations~\cite{Sarma:1996}. Since Fe($3d$) states are overdelocalized in standard DFT, their position [and hence the position of the first two peaks for O($2p$) states] with respect to the valence band maximum turns out to be inaccurate. The third peak in the PDOS for O($2p$) appears at 3.5~eV and it overlaps in energy with La($5d$) and mainly with La($4f$) states, which means that these states hybridize. The broad structure in the range from 4.1 to 7.7~eV of the PDOS for O($2p$) follows closely the profile of the PDOS for La($5d$), which is again a signature of their hybridization. Finally, the remaining peaks in the PDOS for O($2p$) in the range from 7.7~eV upwards correspond to the hybridization between O($2p$) states with Fe($4s$), Fe($4p$), La($6s$), and La($6p$) (not shown). 

Application of the Hubbard $U$ correction to Fe($3d$) and La($4f$) states [see Eq.~\eqref{eq:potU} and Table~\ref{tab:HP}] changes significantly the electronic structure of LFO. In particular, the band gap at the DFT+$U$ level becomes in much better agreement with the experimental value (see Table~\ref{tab:Egap}), though now it is overestimated by 28\%. Note that neglecting the Hubbard $U$ correction for Fe($3d$) states, as was done in DFT+$U$ studies of Ref.~\cite{Nohara:2009}, does not improve the band gap and it was found to be only~0.10~eV (this surprisingly low value is even much smaller than our standard DFT result of 0.65~eV).
Instead, the $GW$ calculations on top of DFT+$U$ of Ref.~\cite{Nohara:2009} give the band gap of 1.78~eV, which underestimates the experimental value by 15\%. As to what concerns the PDOS, as can be seen in Fig.~\ref{fig:PDOS}~(a) (middle panel), Fe($3d$) minority spin states are pushed up in energy when applying the Hubbard $U$ correction to Fe($3d$) states (and, moreover, the $t_{2g}$ and $e_g$ peaks approach each other), and hence the lowest O($2p$) states are also blueshifted. Importantly, the La($4f$) peak is also shifted up in energy due to the Hubbard $U$ correction for these states (see Table~\ref{tab:HP}), and now it overlaps in energy with La($5d$) states, in agreement with calculations of Ref.~\cite{Wu:1997}. Instead, in Ref.~\cite{Nohara:2009} the La($4f$) states appear at too high energies (above 10~eV) both at the DFT+$U$ and $GW$ on top of DFT+$U$ levels of theory, which is due to the application of a too large value of $U$ to these states (7.5~eV). As will be discussed in Sec.~\ref{sec:XANES}, our simulated XANES spectra at the DFT+$U$ level reproduce fairly well the broad feature (in the range 533--539~eV) in the spectrum originating from the hybridized La($4f$) and La($5d$) states, suggesting that our {\it ab initio} predicted position of La($4f$) states is sound. The correct shift of La($4f$) states is crucial for the correct interpretation of the XANES spectra, as will be shown in Sec.~\ref{sec:XANES} (see also Appendix~\ref{secSM:noUonLa4f} for the PDOS when there is no Hubbard $U$ correction for La($4f$) states). Thus, compared to standard DFT, the PDOS for O($2p$) in DFT+$U$ can be divided into three regions: $i)$~a~double-peaked structure due to the overlap with Fe($3d$) states, $ii)$~a~broad structure due to the hybridization with La($5d$) and La($4f$) states, and $iii)$~an~extended structure due to the hybridization with Fe($4s$), Fe($4p$), La($6s$), and La($6p$) states.

The inclusion of the inter-site interactions between Fe($3d$) and O($2p$) states via the Hubbard $V$ correction [see Eq.~\eqref{eq:potV} and Table~\ref{tab:HP}] leads to quantitative changes for the lowest energy empty states. The band gap at the DFT+$U$+$V$ level is increased even further with respect to DFT+$U$ (see Table~\ref{tab:Egap}), and now it overestimates the experimental value by 40\%~\cite{Note:DFTgap}. This latter aspect would need a deeper analysis, which is though not among the objectives of this study. As can be seen in Fig.~\ref{fig:PDOS}~(a) (bottom panel), the PDOS for O($2p$) is modified only slightly. It is important to note that not only did we apply inter-site Hubbard $V$ between Fe($3d$) and O($2p$) states, but also the on-site Hubbard $U$ on Fe($3d$) and La($4f$) are changed by about 0.4~eV and 0.2~eV, respectively, with respect to the DFT+$U$ case: thus the quantitative changes in the PDOS are a combined effect of renormalized $U$ and a finite value of $V$. 

Therefore, in the case of LFO, while the main effect of improving the electronic structure comes from the Hubbard $U$ correction applied to Fe($3d$) and La($4f$) states, the inter-site Hubbard $V$ interactions between Fe($3d$) and O($2p$) states lead to relatively small changes in the lowest empty states. 
However, for the sake of completeness, in Sec.~\ref{sec:XANES} we will present the XANES spectra of LFO computed both on top of DFT+$U$ and DFT+$U$+$V$.

\subsubsection{LaFe$_{0.75}$Ni$_{0.25}$O$_{3}$}
\label{sec:PDOS_LFNO}

LFNO at the DFT level turns out to be metallic, in contradiction to the experimental observation of a semiconducting behavior~\cite{Erat:2009, Idrees:2011}. This failure of standard DFT is due to the same reason which was mentioned for LFO in Sec.~\ref{sec:PDOS_LFO}, i.e. overdelocalization of Fe($3d$) states, but now also overdelocalization of Ni($3d$) states, which both are responsible for the fake metallicity of LFNO. As can be seen in Fig.~\ref{fig:PDOS}~(b) (top panel), the first peak in the PDOS for O($2p$) appears at 0.5~eV and it comes from the hybridization with Ni($3d$) minority spin states and with Fe($3d$) minority spin states, with the latter having much larger intensity (by more than by a factor of 5). Smaller intensity peaks in the PDOS for O($2p$) appear at 1.8 and 2.6~eV, and they originate purely from the hybridization with Fe($3d$) minority spin states. The peak at 3.2~eV is hybridized with La($5d$) and mainly La($4f$), peaks in the range from 3.8 to 7.3~eV are hybridized with La($5d$) states, and finally peaks in the range from 7.3~eV upwards are due to the hybridization with Fe($4s$), Fe($4p$), Ni($4s$), Ni($4p$), La($6s$), and La($6p$) (not shown).

The application of the Hubbard $U$ correction [see Eq.~\eqref{eq:potU} and Table~\ref{tab:HP}] to Fe($3d$), Ni($3d$), and La($4f$) states leads to the opening of a band gap of 0.56~eV (see Table~\ref{tab:Egap}), which is now in agreement with the observed semiconducting behavior of LFNO. As can be seen in Fig.~\ref{fig:PDOS}~(b) (middle panel), the first peak in the PDOS for O($2p$) appears at 0.86~eV and it originates from hybridization uniquely with Ni($3d$) minority spin states. The second peak in the PDOS for O($2p$) has a maximum at 2.6~eV, and it has a weak shoulder around 2.1~eV. It is important to mention that the PDOSs for Fe($3d$) and Ni($3d$) minority spin states both have a double-peak structure, which is due to the $t_{2g}-e_{g}$ splitting; one intense Fe($3d$) peak and one intense Ni($3d$) peak hybridize with O($2p$) and hence responsible for a peak at 2.6~eV with a shoulder at 2.1~eV (see Fig.~\ref{fig:PDOS}~(b), middle panel). The next peak in the PDOS for O($2p$) has a maximum at 3.6~eV and it originates from a hybridization with the Fe($3d$) minority spin states. The broad structure in the PDOS of O($2p$) in the range from 4.1 to 8.0~eV originates from a hybridization with La($5d$) and La($4f$) states, and higher energy peaks have the same origin as was concluded at the level of standard DFT. Similarly to LFO, the blueshift of the La($4f$) states due to the Hubbard $U$ correction turns out to be large and important, as now the first three peaks in the PDOS of O($2p$) are purely due to the hybridization with Ni($3d$) and Fe($3d$) minority spin states with no contribution from La($4f$) states (see Appendix~\ref{secSM:noUonLa4f} for the PDOS when there is no Hubbard $U$ correction for La($4f$) states). Finally, all the remaining high-energy peaks in the PDOS for O($2p$) have the same origin as was described at the DFT level above, with the difference that these structures are blueshifted. 

Inclusion of the inter-site interaction between Fe($3d$) and O($2p$) states and between Ni($3d$) and O($2p$) states via the Hubbard $V$ correction [see Eq.~\eqref{eq:potV} and Table~\ref{tab:HP}] leads to quantitative changes for the PDOS, with a stronger effect than was observed for LFO in Sec.~\ref{sec:PDOS_LFO}. The band gap at the DFT+$U$+$V$ level is increased with respect to DFT+$U$ by almost a factor of 2 (see Table~\ref{tab:Egap}). Unfortunately, to the best of our knowledge, no experimental measurement of this quantity exists for LFNO, so the relevance of our results cannot be assessed. The first peak in the PDOS for O($2p$) is still due to the hybridization uniquely with Ni($3d$) minority spin states, but now this peak appears at 1.3~eV. The next peak, which at the DFT+$U$ level was considered to be a peak with a shoulder, is now split into two clear peaks appearing at 2.1 and 2.9~eV: the former is hybridized with Ni($3d$) minority spin states, while the latter is hybridized with Fe($3d$) minority spin states of larger intensity. The peak at 4.0~eV in the PDOS for O($2p$) is hybridized with Fe($3d$) minority spin states. There are also some changes in the range from 4.0 to 6.0~eV: the PDOS of O($2p$) is more plateau-like with respect to the DFT+$U$ case. All the subsequent peaks remain little affected by switching from DFT+$U$ to DFT+$U$+$V$: they only experience a blueshift, but the shape remains the same.

 \subsubsection{Effect of homovalent substitution of 25\% of Fe atoms by Ni atoms}
\label{sec:PDOS_discussion}

As can be seen in Fig.~\ref{fig:PDOS}, already at the DFT+$U$ level one can make meaningful conclusions about the effect of homovalent substitution of 25\% of Fe atoms by Ni atoms. The first new feature in the PDOS for O($2p$) of LFNO with respect to LFO is that the former contains an extra peak due to the hybridization between O($2p$) states and Ni($3d$) minority spin states (the lowest energy peak). These Ni($3d$) minority spin states appear in the band gap of LFO, and hence the resulting LFNO structure has a band gap which is smaller than the gap of  LFO roughly by a factor of 3--5, according to our DFT+$U$ and DFT+$U$+$V$ calculations. The second difference between the PDOS for O($2p$) between LFNO and LFO is that for the former the nature of the second structure in the PDOS is of a mixed character, namely, it originates not only from the hybridization between O($2p$) states and Fe($3d$) minority spin states, but also due to the hybridization between O($2p$) states and Ni($3d$) minority spin states. The latter hybridization effect [i.e. between O($2p$) and Ni($3d$)] manifests itself either as a shoulder or a clear peak in the PDOS for O($2p$), depending on whether DFT+$U$ or DFT+$U$+$V$ is used. Thus, there is delicate sensitivity of the second structure in the PDOS of LFNO to the inter-site interactions.  
Finally, the peak at around 4~eV is due to the hybridization between O($2p$) and Fe($3d$) minority spin states: this peaks is observed in both LFNO and LFO. Interestingly, in LFNO the splitting of Fe($3d$) states at the DFT+$U$ and DFT+$U$+$V$ levels of theory is larger than in LFO: this is a consequence of changes in the crystal-field and exchange splittings under the substitution of 25\% of Fe atoms by Ni atoms.

\subsection{Oxygen $K$-edge spectra}
\label{sec:XANES}

XANES spectra provide useful information about unoccupied states. In this section we discuss the O $K$-edge XANES spectra of LFO and LFNO as measured in our experiments and as computed using the approach described in Sec.~\ref{sec:comput_approach}. These spectra originate from the transition of an O($1s$) electron to unoccupied $p$ states of O (due to dipole selection rules). The interpretation of all features in the XANES spectra will be made based on the results of the PDOS of Sec.~\ref{sec:PDOS}.

It is important to make a comment about the alignment of the simulated XANES spectra for LFO and LFNO. Inequivalent O atoms can have different core level shifts, which may affect simulated spectra. There are standard procedures to align spectra by taking into account such core level shifts~\cite{England:2011, Stellato:2018}. We have evaluated the core level shifts using a $3 \times 3 \times 2$ supercell for LFO and found that differences between core level shifts for inequivalent O atoms are smaller than 0.02~eV. Hence, core level shifts were neglected for both LFO and LFNO. Moreover, simulated XANES spectra at different levels of theory were aligned using the procedure described in Refs.~\cite{England:2011, Stellato:2018}.

\subsubsection{LaFeO$_3$}
\label{sec:XANES_LFO}

The computed and measured O $K$-edge spectra of LFO are shown in Fig.~\ref{fig:PDOS}~(c). The experimental spectrum is in good agreement with previous measurements~\cite{Abbate:1992, Sarma:1994, Wu:1997, Hayes:2011, Jana:2019}, and it consists of several structures which can be divided into three regions: $i)$~the first two peaks [labeled as 1 and 2] appearing in the range from 529 to 533~eV, $ii)$~a broad structure in the range from 533 to 539~eV, and $iii)$~an extended structure in the range from 539~eV upwards. On the other hand, the theoretical spectrum of LFO depends strongly on the level of theory which is used, and this will be discussed in detail in the following. By comparing Figs.~\ref{fig:PDOS}~(a) and \ref{fig:PDOS}~(c) it is easy to see that the computed O $K$-edge spectra follow very closely the PDOS for O($2p$) at all levels of theory. Therefore, all the discussions about the origin of the peaks in the PDOS for O($2p$) is valid here for the interpretation of the origin of peaks in the O $K$-edge spectrum of LFO. 
From Fig.~\ref{fig:PDOS}~(c) we can see that the XANES spectrum computed using standard DFT is already in quite good agreement with the experiment, however there are some serious drawbacks. More specifically, the first two peaks in XANES from DFT correspond to the experimental peaks 1 and 2, which originate from the hybridization with Fe($3d$) minority spin states. The splitting between these two O($2p$) peaks is 1.48~eV in the experiment of this study (1.2~eV~\cite{Abbate:1992}, 1.6~eV~\cite{Sarma:1994}, 1.4~eV~\cite{Hayes:2011}), while from DFT we obtain 1.64~eV, which is in fairly good agreement with the experiments. As for what concerns the relative intensities of these two peaks, there is no general consensus between the experiments: while in our experimental spectrum the peak~2 has a slightly lower intensity than the peak~1, in Refs.~\cite{Wu:1997,Jana:2019} the peak~2 has a much lower intensity than the peak~1, in Refs.~\cite{Sarma:1994, Hayes:2011} the intensities are approximately equal, and in Ref.~\cite{Abbate:1992} the peak~2 has a larger intensity than the peak~1. Perhaps such variations in the intensities could be related to the nonstoichiometry of the samples~\cite{deGroot:1989}. Our XANES spectrum from DFT shows that the peak~2 has a lower intensity than the peak~1, hence qualitatively in agreement with our measurements and with Refs.~\cite{Wu:1997,Jana:2019}. However, the major problem of XANES from DFT is the third peak [appearing at 532.6~eV in Fig.~\ref{fig:PDOS}~(c)], which is not observed in the experiment. In fact, this peak comes mainly from the hybridization between O($2p$) and La($4f$) states [see Fig.~\ref{fig:PDOS}~(a), top panel], and is thus sensitive to the misplacement of the latters caused by approximate exchange-correlation functionals. The small feature in the experimental spectrum around 532.6~eV is a noise or a consequence of the not perfect stoichiometry of our samples; in Refs.~\cite{Abbate:1992, Sarma:1994, Wu:1997, Hayes:2011, Jana:2019} no such a feature was observed. Therefore, the spurious peak in our XANES simulations on top of DFT at 532.6~eV should not be attributed by any means to this small experimental artifact. The broad structure in XANES from DFT in the range from 533 to 537~eV is in fairly good agreement with the experiment, however the full width at half maximum (FWHM) of this structure is underestimated. Finally, the extended structure in the range from 539~eV upwards is also quite well reproduced in our simulations; however the relative intensities of the peaks in this structure are not the same as in the experiment.

Application of the Hubbard $U$ correction to Fe($3d$) and La($4f$) states alters significantly the XANES spectrum of LFO. As was discussed in Sec.~\ref{sec:PDOS_LFO}, the PDOS of O($2p$) states changed largely due to the shift in energy of the Fe($3d$) and La($4f$) states. The changes in the spectrum  due to such corrections improve (but not everywhere) the consistence between experiment and theory. Namely, the application of the Hubbard $U$ correction to Fe($3d$) states does not in every aspect improve the results, because now the splitting between the $t_{2g}$ and $e_g$ states is much smaller than in DFT, and consequently the distance between the first two peaks in the O $K$-edge is only 0.56~eV, which is underestimated by a factor of 2-3 with respect to experiments. Such a large underestimation of the splitting between first two peaks is responsible for the large redshifting of all the following peaks with respect to the experimental spectrum. Moreover, the relative intensities of these two peaks have also changed, and now the peak~2 is slightly more intense than the peak~1. It is worth noting that our preliminary results for XANES spectra including the core hole on O($1s$) have shown that the latter does not explain the underestimation of the splitting between the first two peaks observed in our DFT+$U$ studies.
On the other hand, in the DFT+$U$ study of Ref.~\cite{Jana:2019} the $t_{2g}$--$e_g$ splitting in Fe is about 1~eV, which is due to the smaller value of the Hubbard $U$ for Fe($3d$) states and other technical details. Instead, in Ref.~\cite{Nohara:2009} the Hubbard $U$ correction for Fe($3d$) states was fully neglected in DFT+$U$ calculations followed by $GW$ calculations, leading to the $t_{2g}$--$e_g$ splitting of about 1.5~eV, though the peak~1 was found to be far too intense with respect to the peak~2 (see Fig.~4~(b) in Ref.~\cite{Nohara:2009}). Thus, our \textit{ab~initio} value of $U = 5.16$~eV for Fe($3d$) states turns out to be too large for the description of the splitting of Fe($3d$) states and consequently the splitting of the first two peaks in the O $K$-edge spectrum, while it is appropriate for the description of, e.g., magnetic moments (see Sec.~\ref{sec:PDOS_LFO}). In the current study we are thus not able to describe accurately all the properties simultaneously: the energetics of valence electrons, magnetic moments, band gaps, and the position of empty electronic states with DFT+$U$ with a global $U$ parameter. Therefore, generalizations beyond the standard DFT+$U$ scheme which is used in this work~\cite{Dudarev:1998} are needed, e.g. by using $m$-resolved $U$ (to have distinct corrections for $t_{2g}$ and $e_g$ states) and/or frequency-dependent extensions. Despite this observation, there is a positive effect from the application of the Hubbard $U$ correction to La($4f$) states which are pushed up in energy and hence overlap with La($5d$) states, as was discussed in Sec.~\ref{sec:PDOS_LFO}. Thanks to this, there is no longer the spurious peak of the DFT case, instead there is a broader structure in the range from 532 to 536~eV, which is in better agreement with the experimental XANES. As was pointed out above, such a broad structure overall has redshifted with respect to the DFT one, thus worsening slightly the agreement with the experiment. As shown in Appendix~\ref{secSM:noUonLa4f}, the neglect of Hubbard~$U$ on La($4f$) states has a dramatic effect on the resulting XANES spectra: the comparison with the experiment is misleading and the interpretation of the origin of the peaks is confusing even at the DFT+$U$ level with $U$ only on Fe($3d$) states. 

It turns out that inter-site Hubbard $V$ interactions between Fe($3d$) and O($2p$) states change only slightly the XANES spectrum, in agreement with the findings in Sec.~\ref{sec:PDOS_LFO}. The only change with respect to the DFT+$U$ case is a further redshift of the broad structure which comes after the peaks~1 and 2. This suggests that the Hubbard~$V$ corrections in LFO are not crucial, and hence DFT+$U$ is responsible for the main effect. However, this is likely not a general trend of ABO$_3$ pristine perovskites, and hence the effect of inter-site $V$ on XANES spectra should be carefully investigated case-by-case, paying special attention to the covalency of the system under study.

\subsubsection{LaFe$_{0.75}$Ni$_{0.25}$O$_{3}$}
\label{sec:XANES_LFNO}

The computed and measured O $K$-edge spectra of LFNO are shown in Fig.~\ref{fig:PDOS}~(d).  The experimental spectrum is in good agreement with measurements for LaFe$_{0.7}$Ni$_{0.3}$O$_3$~\cite{Sarma:1994}, and it consists of several structures which can be divided into three regions (similarly to LFO): $i)$~the first three peaks [labeled as 1$^\prime$, 1, and 2] appearing in the range from 528 to 533~eV, $ii)$~a broad structure in the range from 533 to 538~eV, and $iii)$~an extended structure in the range from 538~eV upwards. The experimental spectrum of LFNO resembles closely the one of LFO, but of course there are important differences. In particular, there is a new pre-peak (peak~1$^\prime$), and the intensities of peaks~1 and 2 are reduced in LFNO with respect to LFO, while the other structures in the spectra of these two materials are essentially the same.  On the other hand, the theoretical spectrum of LFNO depends strongly on the level of theory which is used, similarly to LFO, and this will be discussed in detail in the following. By comparing Fig.~\ref{fig:PDOS}~(b) with Fig.~\ref{fig:PDOS}~(d) we can easily see that the computed O $K$-edge spectra follow very closely the PDOS for O($2p$) at all levels of theory. Therefore, all the discussions about the origin of the peaks in the PDOS for O($2p$) is valid here for the interpretation of the origin of peaks in the O $K$-edge spectrum of LFNO. 

At the DFT level, there is seemingly quite good agreement between theory and experiment, apart from a too large intensity of the peak~1$^\prime$. However, we stress that this is only a misleading impression, because LFNO is metallic at the DFT level [see Fig.~\ref{fig:PDOS}~(b), top panel]. In practice, to plot the XANES spectrum at the DFT level, we cut off the occupied states and applied a smearing to smoothen the spectrum, which then looks outwardly quite good. But in reality the peak~1$^\prime$ in DFT is completely wrong. In fact, based on the results of Sec.~\ref{sec:PDOS_LFNO}, the ``fake" peak~1$^\prime$ originates from hybridizations between O($2p$) states with mainly Fe($3d$) and to a much smaller extent with Ni($3d$) minority spin states, which both contribute to the density of states at the Fermi level. For the same reason the attribution of the peaks~1 and 2 in the XANES from DFT to the peaks~1 and 2 in the experimental XANES of LFNO is also very problematic. For example, the peak~1 from DFT turns out to be due to the hybridization between O($2p$) and Fe($3d$) minority spin states, which is true only partially (as will be seen in the following). Peak~2 from DFT is instead completely wrong, as it is seemingly due to mainly the hybridization between O($2p$) and La($4f$) states. This is not surprising because already in LFO we saw that La($4f$) states create serious problems at the DFT level. The broad structure in XANES from DFT in the range from 532 to 536~eV is due to the hybridization between O($2p$) and La($5d$), as was shown in Sec.~\ref{sec:PDOS_LFNO}, which is also partially correct (as will be shown in the following, La($4f$) states also appear in this energy range). However, the position, shape, and FWHM of this structure at the DFT level is quantitatively not perfect. Finally, the extended structure from 538~eV upwards in DFT corresponds to the extended structure in experimental XANES, and its origin was discussed in Sec.~\ref{sec:PDOS_LFNO}.

Application of the Hubbard $U$ correction to Fe($3d$), Ni($3d$), and La($4f$) states changes significantly the XANES spectrum of LFNO. As was discussed in Sec.~\ref{sec:PDOS_LFNO}, the PDOS of O($2p$) states changed largely due to the shift in energy of the Fe($3d$), Ni($3d$), and La($4f$) states. Therefore, we want to stress that thanks to the inclusion of the Hubbard $U$ corrections for Fe($3d$) and Ni($3d$) states the system goes from metal to semiconductor, which highlights the crucial role of Hubbard corrections for these states and which thus need to be included in simulations of the XANES spectra of LFNO. In DFT+$U$, the peak~1$^\prime$ originates from the hybridization between O($2p$) states and Ni($3d$) minority spin states. This is not surprising, because in LFO there are no Ni atoms and hence no a pre-peak. Peak~1 is due to the hybridization of O($2p$) states with Fe($3d$) and Ni($3d$) minority spin states (this peak has a maximum at higher energies and a small shoulder at smaller energies), and peak~2 is due to the hybridization with Fe($3d$) minority spin states only, as was discussed in Sec.~\ref{sec:PDOS_LFNO}. Therefore, the shift of La($4f$) states due to the Hubbard $U$ correction turns out to be very important, similarly to the case of LFO (see Appendix~\ref{secSM:noUonLa4f} for the case when the Hubbard $U$ correction is not applied to La($4f$) states). Thus, qualitatively XANES from DFT+$U$ reproduces the peaks 1$^\prime$, 1, and 2, but the relative intensities of these peaks and the positions of peaks~1 and 2 do not match exactly those in the experimental spectrum.
The broad structure in the range from 532 to 536~eV has slightly changed its shape (due to the overlap in energy of La($4f$) and La($5d$) states) and slightly blueshifted, and the extended structure in the range from 538~eV upwards does not show any significant changes with respect to the standard DFT case. 

Finally, at the DFT+$U$+$V$ level the agreement between the computed XANES spectrum and the experimental one is also qualitatively fairly good (similarly to DFT+$U$). The main difference with respect to DFT+$U$ is that the peak~1 is split into two peaks, as was also observed in the PDOS for LFNO [see Fig.~\ref{fig:PDOS}~(b)]. Such a splitting is due to the slightly larger energy separation between Ni($3d$) and Fe($3d$) minority spin states, which occurs due to the inclusion of the Hubbard $V$ interactions of these states with O($2p$) states and slight changes in the values of $U$ (see Table~\ref{tab:HP}). Peak~2 has a larger overlap with the broad structure which appears at larger energies due to the redshift of the latter after the inclusion of Hubbard $V$. Such a redshift of the broad structure after the peak~2 worsens the agreement with the experiment. Importantly, the relative intensities of the peaks~1 and 2 are in closer agreement with the experiment at the DFT+$U$+$V$ level than at the DFT+$U$ one. 

Overall, our simulations at the DFT+$U$+$V$ level capture the main features in the XANES spectrum of LFNO and allow us to interpret the origin of all peaks. In particular, our calculations show that the replacement of 25\% of Fe with Ni leads to a change of character of the bottom of the low-energy conduction bands, as in the Ni-doped material it is formed by hybridized O($2p$)--Ni($3d$) minority spin states (which confirms the conclusions made in Ref.~\cite{Idrees:2011} that are purely based on experimental findings) -- an effect completely missing in simulations performed using standard gradient-corrected functionals. 

\begin{figure*}[t]
\begin{center}
 \subfigure[]{\includegraphics[width=0.48\linewidth]{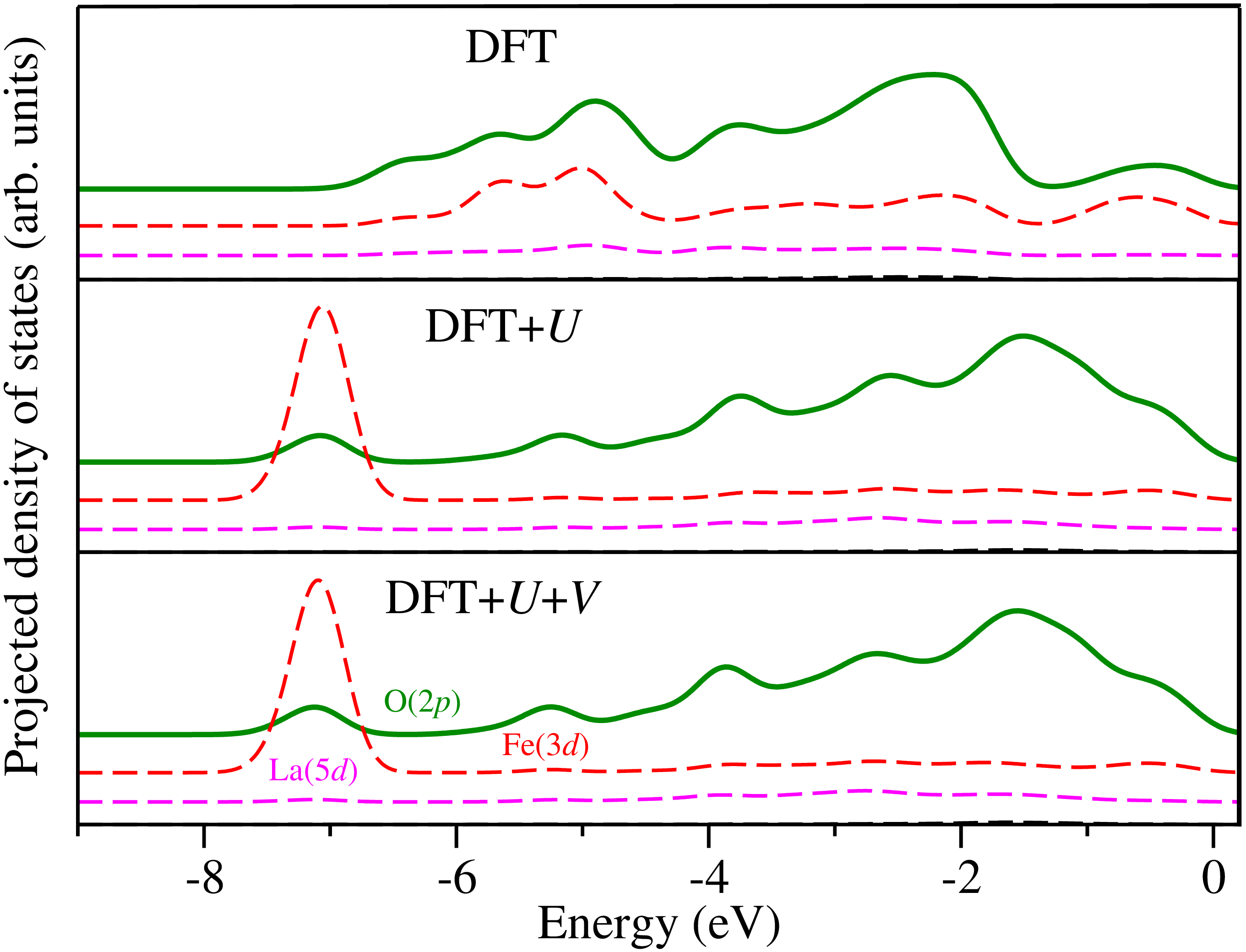}\hspace{0.1cm}}
 \subfigure[]{\includegraphics[width=0.48\linewidth]{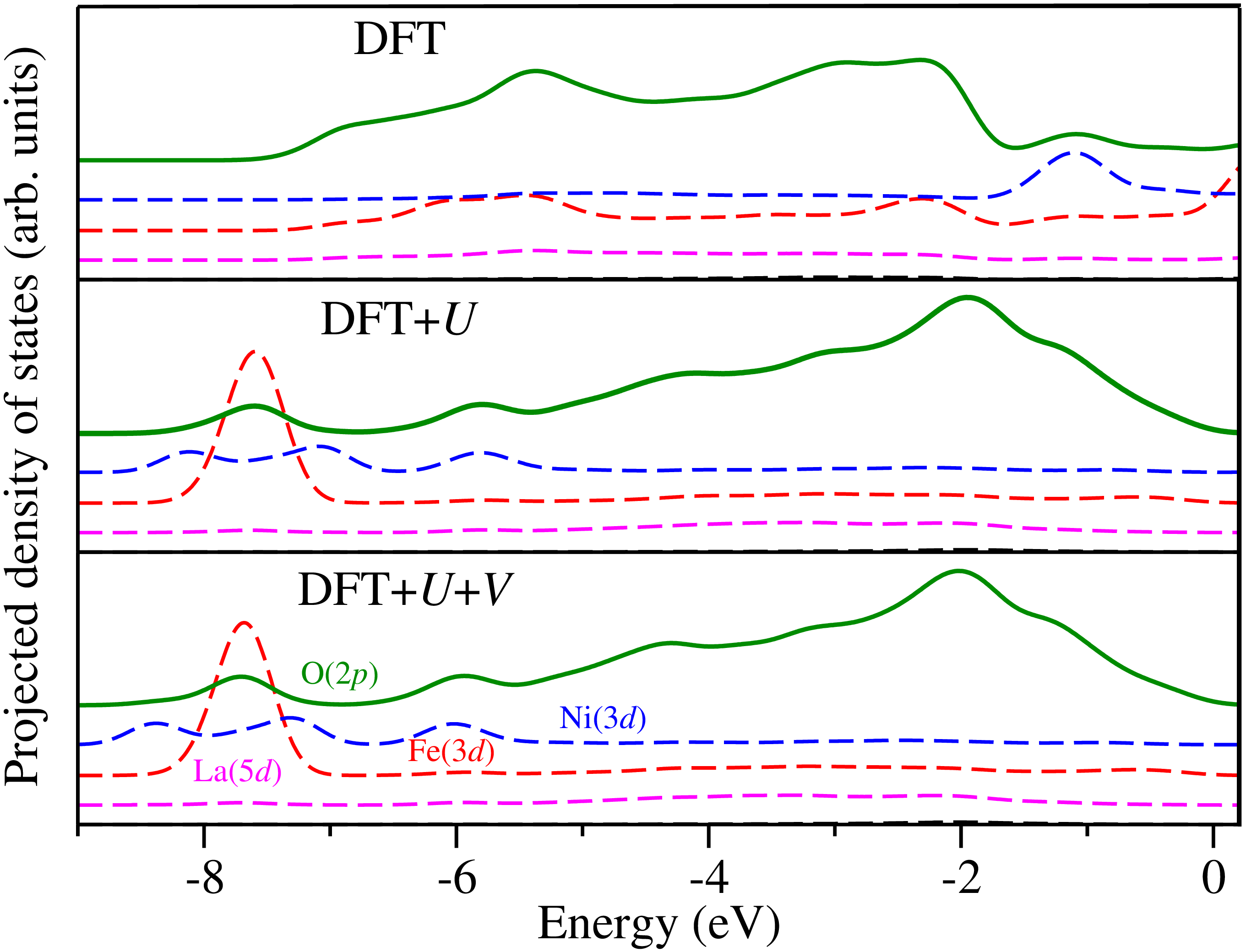}}
 \caption{Projected density of occupied states computed using DFT, DFT+$U$, and
DFT+$U$+$V$ for (a)~LaFeO$_3$ and (b)~LaFe$_{0.75}$Ni$_{0.25}$O$_{3}$, using Hubbard parameters listed in Table~\ref{tab:HP}. The zero of energy corresponds to the valence band maximum (or Fermi energy in the case of LaFe$_{0.75}$Ni$_{0.25}$O$_{3}$ at the DFT level of theory, which comes out to be metallic). PDOS components were shifted vertically for the sake of clarity. The PDOSs for Fe($3d$) and Ni($3d$) states show only the majority spin components on all panels (the minority spin components correspond to unoccupied states).}
\label{fig:PDOS_valence}
\end{center}
\end{figure*}

\section{Conclusions}
\label{sec:Conclusions}

We have presented a joint theoretical and experimental study of the O $K$-edge spectra of LFO and LFNO using XANES measurements and comparing it with DFT calculations using extended Hubbard functionals. Thanks to the inclusion of the on-site $U$ and inter-site $V$ Hubbard corrections, which were determined from first-principles (no adjustable or fitting parameters), we were able to accurately attribute the origin of all pre-edge peaks measured in the O $K$-edge XANES spectra of LFO and LFNO. 

From a methodological point of view, we generalized the method of Ref.~\cite{Gougoussis:2009} to the case of extended Hubbard functionals, i.e. the DFT+$U$+$V$ approach. The inclusion of both on-site and inter-site Hubbard interactions allows us to obtain an accuracy for the low-energy features of XANES spectra comparable to that of hybrid functionals but at substantially lower computational cost~\cite{TancogneDejean:2019, Ricca:2020}.

We demonstrated that the application of the on-site Hubbard $U$ correction to Fe($3d$), Ni($3d$), and La($4f$) states and inter-site $V$ corrections between these states and O($2p$) improves the description of the electronic structure of LFO and LFNO with respect to gradient-corrected functionals. In particular, we found that the O $K$-edge pre-peak (peak~1$^\prime$) in LFNO measures the conduction band bottom and it is due to the hybridization between O($2p$) states and Ni($3d$) minority spin states. Remarkably, the use of gradient-corrected functionals with no Hubbard corrections fails in reproducing both the semiconducting character of Ni-substituted LFO and the pre-edge features of O $K$-edge XANES spectra.

Our work outlines the crucial role of Hubbard interactions to describe the electronic structure of complex transition-metal oxides and suggests that further  developments of extended Hubbard functionals, such as generalization to $m$-dependent Hubbard interactions and, possibly, inclusion of  dynamical effects, could lead to even more accurate modeling of spectroscopic properties of TM oxides. The XANES simulation method based on DFT+$U$+$V$ is now publicly available to the community at large~\cite{QuantumESPRESSO:Gitlab, QuantumESPRESSO:website}. It would be interesting to see other studies on the influence of inter-site Hubbard $V$ corrections on other classes of complex materials by employing the tool presented here and comparing with XANES spectra, and how these translate in differences in the functionality of devices.

In the particular case considered here, experiments do not allow us to quantitatively discriminate between the results with the on-site term alone or both on-site and inter-site terms. Still, there is a non-negligible effect of the latter inter-site Hubbard terms both on the ground-state properties (electronic band gap) and on the XANES spectral shape, with reasonable trends. This shows that the proposed theoretical approach is ready to be applied for interpretations of future, accurate spectroscopy experiments on complex TM compounds. Finally, the influence of inter-site Hubbard interactions on XANES spectra is expected to be even of greater importance in mixed-valence TM compounds~\cite{Cococcioni:2019}. In many of these systems the on-site Hubbard $U$ correction alone fails to accurately predict the electronic occupations and energies when TM elements with different oxidation states are present simultaneously, and where inter-site $V$ is truly fundamental to restore the correct balance between localization of states and their hybridization. Moreover, whenever perovskite materials contain oxygen vacancies or defects, once again inter-site Hubbard interactions are absolutely necessary to describe accurately the formation energies and energetics in general~\cite{Ricca:2020}, and which thus are expected to be of crucial importance for calculations of XANES spectra. Therefore, further applications of the novel XANES method based on DFT+$U$+$V$ are envisaged to clarify further the impact of inter-site interactions on XANES spectra of complex TM compounds.

The data used to produce the results of this work are available on the Materials Cloud Archive~\cite{MaterialsCloudData}.

\begin{figure*}[t]
\begin{center}
 \subfigure[]{\includegraphics[width=0.48\linewidth]{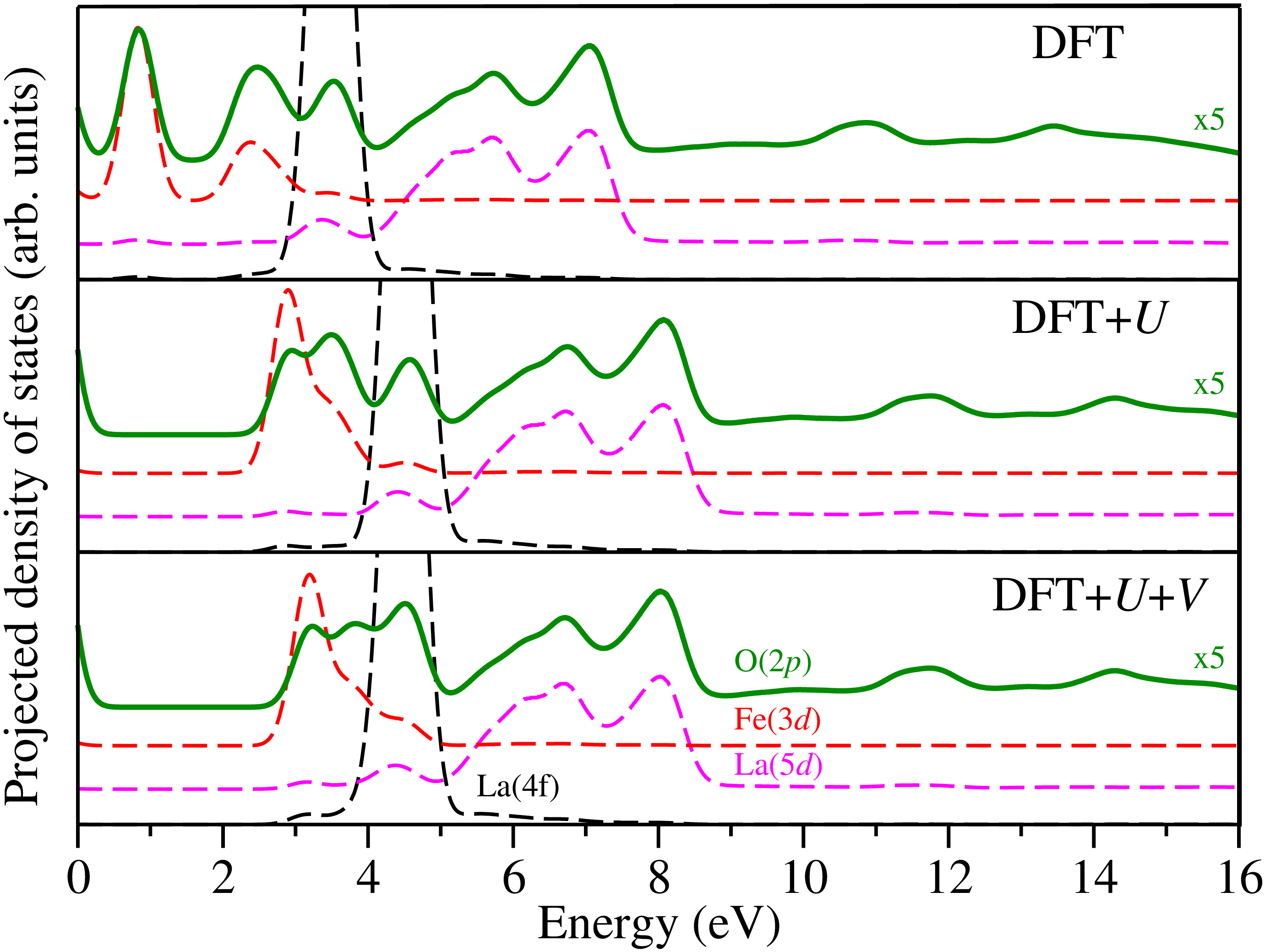}\hspace{0.05cm}}
 \subfigure[]{\includegraphics[width=0.48\linewidth]{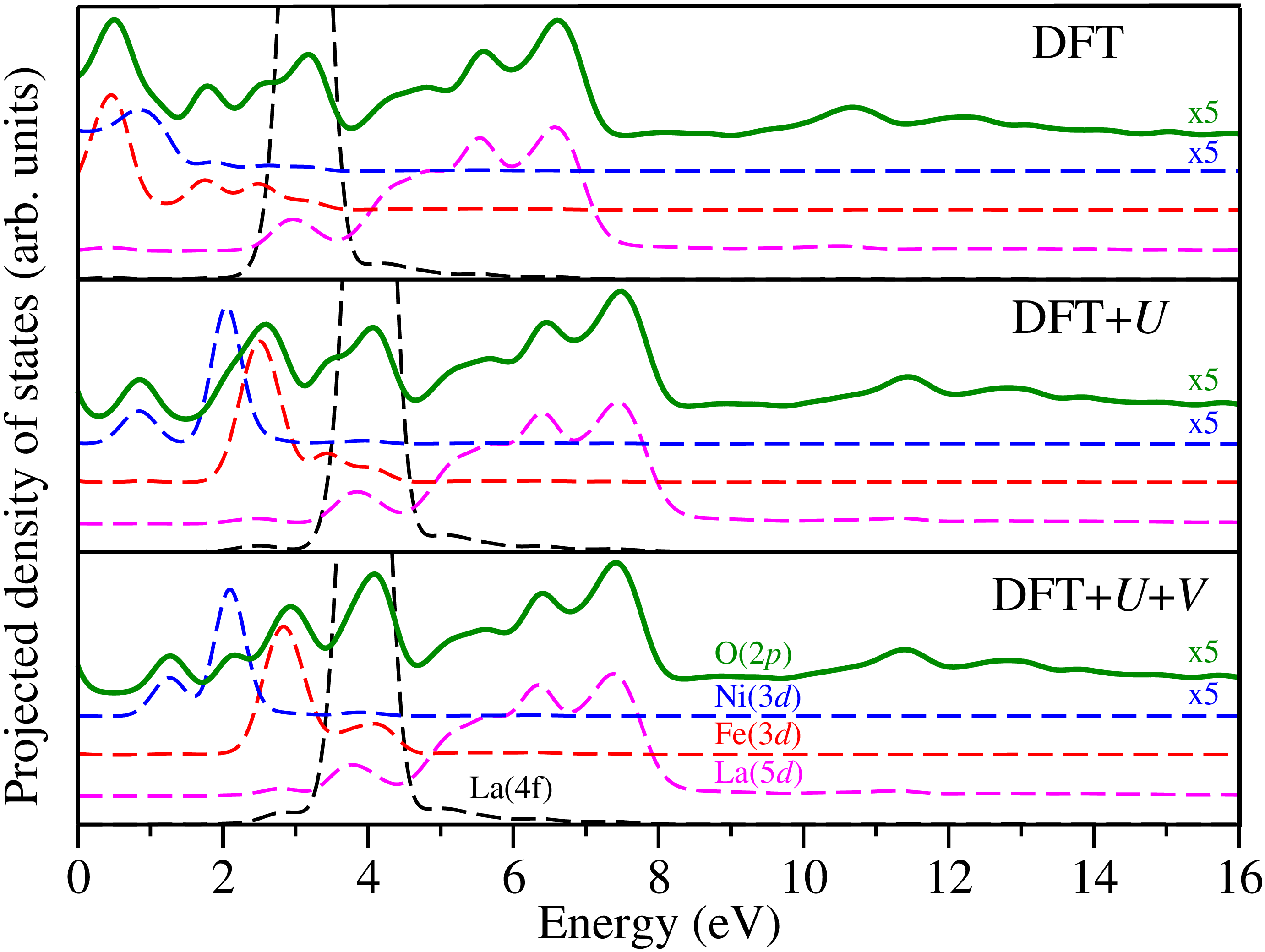}}
 \subfigure[]{\includegraphics[width=0.48\textwidth]{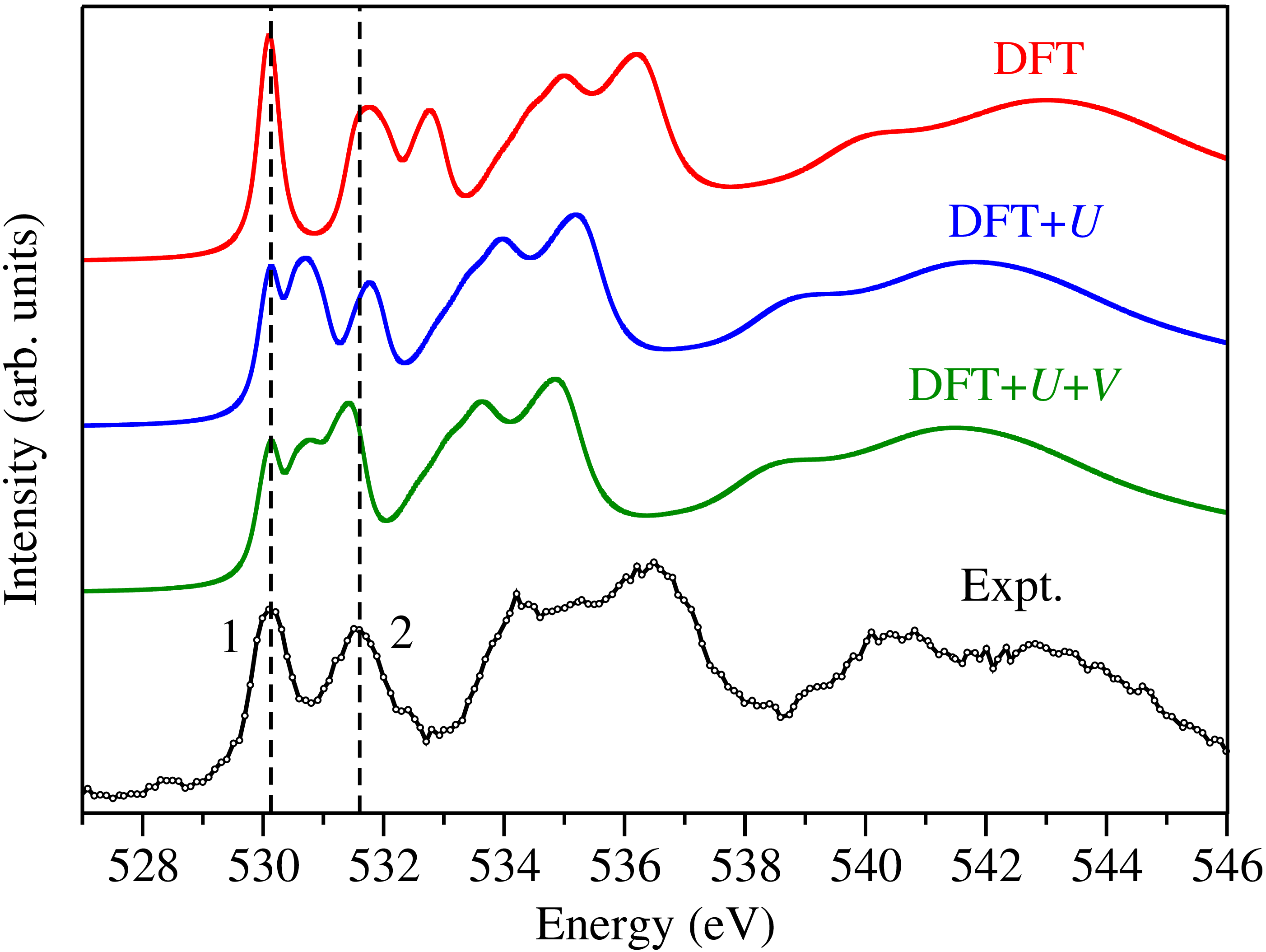}\hspace{0.05cm}}
 \subfigure[]{\includegraphics[width=0.48\textwidth]{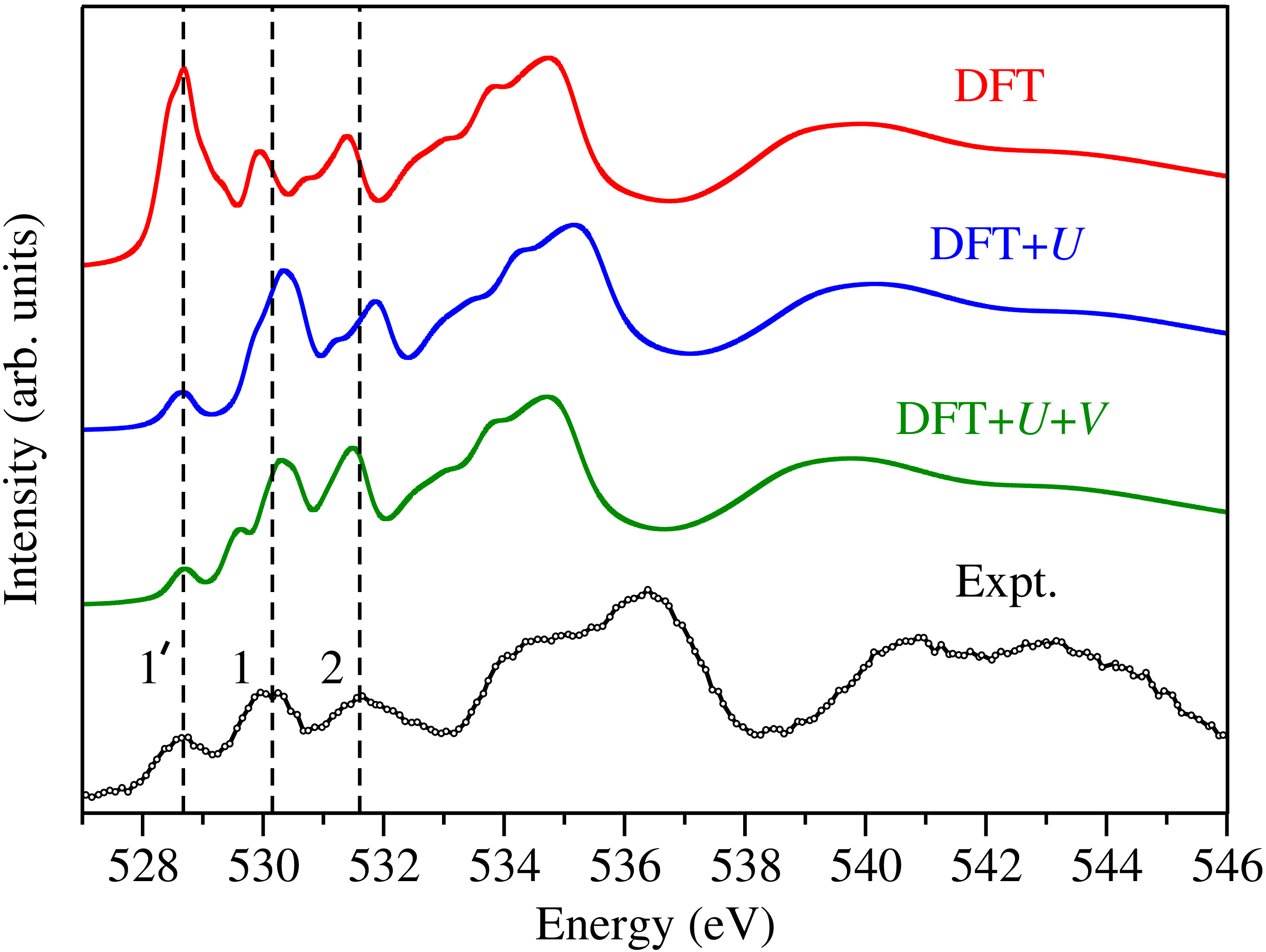}}
\caption{Projected density of empty states computed using DFT, DFT+$U$, and
DFT+$U$+$V$ for (a)~LaFeO$_3$, and (b)~LaFe$_{0.75}$Ni$_{0.25}$O$_{3}$, using Hubbard parameters listed in Table~\ref{tabSM:HP}. The zero of energy corresponds to the valence band maximum (or Fermi energy in the case of LaFe$_{0.75}$Ni$_{0.25}$O$_{3}$ at the DFT level of theory, which comes out to be metallic). The PDOSs for O($2p$) and Ni($3d$) states were multiplied by a factor of 5 for easier comparison with other PDOS components; the PDOSs for Fe($3d$) and Ni($3d$) states show only the minority spin components on all panels since the majority spin components correspond to fully occupied states. Comparison between the experimental and theoretical oxygen $K$-edge XANES spectra of (c)~LaFeO$_3$, and (d)~LaFe$_{0.75}$Ni$_{0.25}$O$_3$. The energy scales for the experimental spectra of LFO and LFNO are aligned (by a rigid shift) in order to facilitate the comparison. PDOS and XANES spectra are shifted vertically for clarity.}
\label{figSM:PDOS_XANES}
\end{center}
\end{figure*}

\section*{ACKNOWLEDGMENTS}

This research was partially supported by the European Commission FP6 Marie Curie project HiTempEchem No.~042095, the European Commission (EU FP6 MIRG 042095), the Swiss-China project IP03-092008, the Swiss National Science Foundation (SNSF) through Grants No.~116688 and No.~179138 and its National Centre of Competence in Research (NCCR) MARVEL. X.Z. acknowledges funding from the National Natural Science Foundation of China under Grant No.~51002130. I.T., M.C., and N.M. acknowledge partial support from the EU-H2020 research and innovation programme under Grant Agreement No.~654360 within the framework of the NFFA Europe Transnational Access Activity. Computer time was provided by CSCS (Piz Daint) through Project No.~s836. This research used resources of the Advanced Light Source (Beamline~9.3.2), a U.S. DOE Office of Science User Facility under contract No.~DE-AC02-05CH11231.

\appendix

\section{Projected density of occupied states}
\label{secSM:PDOS_valence}

Figure~\ref{fig:PDOS_valence} shows the PDOS for occupied states of LFO and LFNO computed at three levels of theory: DFT, DFT+$U$, and DFT+$U$+$V$. Overall it can be seen that the inclusion of Hubbard corrections changes dramatically the shape of the PDOS. In particular, Fe($3d$) majority spin states are very delocalized in standard DFT, while they become very localized at the level of DFT+$U$ and DFT+$U$+$V$ and the corresponding peaks in the PDOS are sharp and are shifted to lower energies. Consequently, the shape of the O($2p$) states is also changed largely in both LFO and LFNO. The PDOS for Ni($3d$) majority spin states of LFNO are also changed significantly due to a better description of their localized nature in the Hubbard-corrected DFT, and the corresponding peaks in the PDOS are also shifted to lower energies. For both materials, the difference between DFT+$U$ and DFT+$U$+$V$ results is quite small but noticeable: the positions of peaks for Fe($3d$) and Ni($3d$) states vary by a fraction of eV. We are not aware of any high-resolution photoemission data for LFO and LFNO in the extended energy range, thus we cannot make any conclusions about the accuracy of DFT+$U$ and DFT+$U$+$V$ predictions for the valence states in these materials. 

Finally, as was also observed experimentally~\cite{Sarma:1994}, the introduction of Ni in LFO does not affect markedly the oxidation state of oxygen. L\"owdin charges of the oxygen atoms are only 0.1 electrons/atom different upon adding Ni with respect to the LFO case (DFT), with this difference decreasing to 0.03 electrons/atom for DFT+$U$ and DFT+$U$+$V$.

\section{Effect of neglecting Hubbard $U$ on La($4f$) states}
\label{secSM:noUonLa4f}

In this appendix we discuss the effect of disregarding the Hubbard $U$ correction for La($4f$) states in LFO and LFNO, i.e., these states are treated at the standard DFT level with the GGA functional.

First of all, we recomputed the Hubbard parameters for LFO and LFNO using the method described in Sec.~\ref{sec:comput_approach}, within both DFT+$U$ and DFT+$U$+$V$ approaches. The results are shown in Table~\ref{tabSM:HP}. By comparing this table with Table~\ref{tab:HP}, we can see that the changes in values of $U$ and $V$ are extremely small. This finding demonstrates that the contribution of La($4f$) states to the screening of electronic interactions is negligible. This is not surprising since La($4f$) states are empty. Using the Hubbard parameters listed in Table~\ref{tabSM:HP}, we recomputed the PDOS and XANES spectra for LFO and LFNO, and the results are shown in Fig.~\ref{figSM:PDOS_XANES}. 

\begin{table*}[t]
 \begin{center}
  \begin{tabular}{lccccccc}
    \hline\hline
                  &                \multicolumn{2}{c}{LaFeO$_3$}                      & \phantom{a} &                                        \multicolumn{4}{c}{LaFe$_{0.75}$Ni$_{0.25}$O$_{3}$}                              \\ \cline{2-3} \cline{5-8}
                  &  $U_{\mathrm{Fe}(3d)}$ &  $V_{\mathrm{Fe}(3d) - \mathrm{O}(2p)}$  &  &$U_{\mathrm{Fe}(3d)}$  &  $U_{\mathrm{Ni}(3d)}$ & $V_{\mathrm{Fe}(3d) - \mathrm{O}(2p)}$ & $V_{\mathrm{Ni}(3d) - \mathrm{O}(2p)}$   \\ \hline
DFT+$U$           &         5.17           &                  --                      &  &    5.28/5.36         &         7.23           &             --                          &               --                        \\  
DFT+$U$+$V$       &         5.54           &              0.76--0.79                  &  &    5.75/5.84         &         7.66           &          0.64--1.01                     &           0.68--0.99                    \\ \hline\hline
   \end{tabular}
 \end{center}
\caption{Self-consistent Hubbard parameters (in eV) for LaFeO$_3$ and LaFe$_{0.75}$Ni$_{0.25}$O$_{3}$ computed using DFPT~\cite{Timrov:2018, Timrov:2020} when using pseudopotentials from the Pslibrary~0.3.1 and 1.0.0~\cite{Dalcorso:2014, Note:Pseudopotentials}. In the case of LaFe$_{0.75}$Ni$_{0.25}$O$_{3}$ there are two inequivalent Fe sites, thus two values of $U$ for Fe($3d$) are specified. For both materials, the inter-site $V$ is specified as a range of values, because there are various inequivalent pairs of neighbors.}
\label{tabSM:HP}
\end{table*}

As can be seen in Fig.~\ref{figSM:PDOS_XANES}~(a), at the DFT+$U$ and DFT+$U$+$V$ levels of theory, the position of the La($4f$) states is incorrect because these states are treated using GGA. Similarly to the standard DFT [see Fig.~\ref{figSM:PDOS_XANES}~(a), top panel] the La($4f$) states overlap very little with the La($5d$) states; instead, La($4f$) states are located very close to the Fe($3d$) minority spin states. As a consequence, by looking at the O $K$-edge XANES spectra of LFO in Fig.~\ref{figSM:PDOS_XANES}~(c) we can see that a spurious peak just above the peak~2 (i.e., around 531-532~eV) is present even at the DFT+$U$ and DFT+$U$+$V$ levels of theory, in contradiction to experimental findings.

In the case of LFNO, the effect of La($4f$) states treated at the GGA level is similar to that of the case of LFO. More specifically, as can be seen in Fig.~\ref{figSM:PDOS_XANES}~(b), at the DFT+$U$ and DFT+$U$+$V$ levels of theory, La($4f$) states overlap very little with La($5d$) states, and instead they appear in the energy interval where Fe($3d$) minority spin states are located. This creates difficulties for the interpretation of peaks in the XANES in Fig.~\ref{figSM:PDOS_XANES}~(d). Indeed, it seems that the peak~2 is in quite good agreement with the experimental peak~2. However, the origin of this peak has nothing to do with the hybridization between O($2p$) and La($4f$) states at the energy of peak~2, as discussed in Sec.~\ref{sec:XANES_LFNO}. 

%\bibliography{references.bib}

\begin{thebibliography}{106}
\bibitem [{Not({\natexlab{c}})}]{Note:DFTgap}%
  \BibitemOpen
  \href@noop {} {} \bibinfo {note} {{It is worth mentioning
  that neither DFT+$U$ nor DFT+$U$+$V$ are devised to give accurate band gaps
  of materials, though often the computed gaps are in better agreement with the
  experimental values (with respect to standard DFT).}}\BibitemShut {Stop}%
\end{thebibliography}

%

\end{document}